\documentclass[%
superscriptaddress,
reprint,
nofootinbib,
 amsmath,amssymb,
 aps, physrev,
]{revtex4-2}
\usepackage{orcidlink}
\usepackage{booktabs}
\usepackage{float}
\usepackage{subfig}
\usepackage{graphicx}% Include figure files
\usepackage{subcaption}   % 用于创建子图
\usepackage{rotating}  % for 
\usepackage{bm}% bold math
\usepackage{hyperref}% add hypertext capabilities
\usepackage{tabularx}   % 用于创建能自动调整列宽的表格
\usepackage{amsmath}    % 推荐，用于在数学公式中插入文本
\usepackage{graphicx}% Include figure files
\usepackage{dcolumn}% Align table columns on decimal point
\usepackage{xcolor}
\begin{document}

%\preprint{Manuscript submitted to Physics Letters B}

\title{Application of interpretable data-driven methods for the reconstruction of supernova neutrino energy spectra following fast neutrino flavor conversions}

% Force line breaks with \\

\author{Haihao~Shi\orcidlink{0009-0007-9418-2632}}\thanks{Haihao Shi and Zhenyang Huang contribute equally to this work}
 \email{shihaihao@xao.ac.cn}
%\altaffiliation[]{}%Lines break automatically or can be forced with \\
%\altaffiliation[]{}
\affiliation{School of Physical Science and Technology, Xinjiang University, Urumqi 830046, People’s Republic of China}

\affiliation{Xinjiang Astronomical Observatory, Chinese Academy of Sciences, Urumqi 830011, People’s Republic of China}

\affiliation{
School of Astronomy and Space Science, University of Chinese Academy of Sciences, Beijing 101408, People’s Republic of China}

\affiliation{International Centre of Supernovae, Yunnan Key Laboratory of Supernova Research, Yunnan Observatories, Chinese Academy of Sciences, Kunming 650216, People’s Republic of China}

\affiliation{Key Laboratory for Structure and Evolution of Celestial Objects, Chinese Academy of Sciences, Kunming 650216, People’s Republic of China}

\author{Zhenyang~Huang\orcidlink{0009-0009-8188-5632}}
 \email{huangzhenyang@xao.ac.cn}
\affiliation{Xinjiang Astronomical Observatory, Chinese Academy of Sciences, Urumqi 830011, People’s Republic of China}
\affiliation{
School of Astronomy and Space Science, University of Chinese Academy of Sciences, Beijing 101408, People’s Republic of China}

\author{Qiyu~Yan\orcidlink{0009-0001-0529-1172}}%
 \email{yanqiyu@st.gxu.edu.cn}
\affiliation{Guangxi Key Laboratory for Relativistic Astrophysics, School of Physical Science and Technology, Guangxi University, Nanning
530004, People’s Republic of China}

\author{Junda~Zhou\orcidlink{0009-0004-7473-1727}}%
 \email{zhoujunda@ynao.ac.cn}
\affiliation{International Centre of Supernovae, Yunnan Key Laboratory of Supernova Research, Yunnan Observatories, Chinese Academy of Sciences, Kunming 650216, People’s Republic of China}

\affiliation{Key Laboratory for Structure and Evolution of Celestial Objects, Chinese Academy of Sciences, Kunming 650216, People’s Republic of China}
\affiliation{
School of Astronomy and Space Science, University of Chinese Academy of Sciences, Beijing 101408, People’s Republic of China}

\author{Guoliang~L\"u\orcidlink{0000-0002-3839-4864} }
\thanks{Corresponding author}
\email{guolianglv@xao.ac.cn}

\affiliation{School of Physical Science and Technology, Xinjiang University, Urumqi 830046, People’s Republic of China}
\affiliation{Xinjiang Astronomical Observatory, Chinese Academy of Sciences, Urumqi 830011, People’s Republic of China}

\author{Xuefei~Chen \orcidlink{0000-0001-5284-8001}}%
\thanks{Corresponding author}
 \email{cxf@ynao.ac.cn}
 
\affiliation{International Centre of Supernovae, Yunnan Key Laboratory of Supernova Research, Yunnan Observatories, Chinese Academy of Sciences, Kunming 650216, People’s Republic of China}

\affiliation{Key Laboratory for Structure and Evolution of Celestial Objects, Chinese Academy of Sciences, Kunming 650216, People’s Republic of China}

\affiliation{
School of Astronomy and Space Science, University of Chinese Academy of Sciences, Beijing 101408, People’s Republic of China}

\date{\today}% It is always \today, today,

\begin{abstract}

% Neutrinos can experience fast flavor conversions (FFCs) in highly dense astrophysical environments, such as core-collapse supernovae and neutron star mergers. 
Neutrinos can experience fast flavor conversions (FFCs) in highly dense astrophysical environments, such as core-collapse supernovae and neutron star mergers, potentially affecting energy transport and other processes.
The simulation of fast flavor conversions under realistic astrophysical conditions requires substantial computational resources and involves significant analytical challenges.
While machine learning methods like Multilayer Perceptrons have been used to accurately predict the asymptotic outcomes of FFCs, their \lq black-box\rq \ nature limits the extraction of direct physical insight. 
% To \textcolor{pink}{address} 
To \textcolor{black}{mitigate} this limitation, we employ two distinct interpretable machine learning frameworks—Kolmogorov-Arnold Networks (KANs) and Sparse Identification of Nonlinear Dynamics (SINDy)—to \textcolor{black}{learn interpretable surrogates for the asymptotic input--output mapping} from a  FFC simulation dataset.
Our analysis reveals a fundamental trade-off between predictive accuracy and model simplicity. 
The KANs demonstrates high fidelity in reconstructing post-conversion neutrino energy spectra, achieving accuracies of up to 90\%. 
% \textcolor{pink}{In contrast, SINDy uncovers a remarkably concise, low-rank set of governing equations, offering maximum interpretability but with lower predictive accuracy. } 
\textcolor{black}{In contrast, SINDy yields a low rank, compact closed-form approximation of the input–output mapping, at the expense of some predictive accuracy.}
Critically, \textcolor{black}{using these structured and sparse surrogates as diagnostic tools,} we identify that the system's evolution is most sensitive to the initial number density of heavy-lepton neutrinos when FCCs are triggered, compared to other physical quantities.
Ultimately, this work provides a methodological framework for interpretable machine learning that supports genuine data-driven \textcolor{black}{scientific} discovery in astronomy and astrophysics, going beyond prediction alone.

\end{abstract}

\maketitle

%\tableofcontents

\section{Introduction}
Neutrino emission plays a crucial role in the dynamics of core-collapse supernovae (CCSNe) and neutron star mergers (NSMs) \cite{Janka:2006fh,Burrows:2020qrp, Janka:2012wk, Foucart:2022bth, Kyutoku:2021icp, Colgate:1966ax, Lattimer:1974slx,Johns:2025mlm,2025arXiv250519592X}, two astrophysical environments characterized by extreme densities and energies. Under these conditions, neutrinos undergo complex flavor conversion processes due to coherent forward scattering interactions within dense background neutrino gases \cite{pantaleone:1992eq, sigl1993general, Pastor:2002we,duan:2006an, duan:2006jv, duan:2010bg, Mirizzi:2015eza, volpe2023neutrinos}.

A significant recent advancement in this area is the discovery of fast flavor conversions (FFCs), which occur over remarkably short spatial scales in the vacuum \cite{Sawyer:2005jk, Sawyer:2015dsa,
Chakraborty:2016lct, Wu:2017qpc,Xiong:2021dex,George:2024zxz,Wang:2025nii}. A necessary and sufficient condition for the occurrence of FFC is that the angular distribution of the neutrino lepton number, defined as:
\begin{equation}\label{eq1}
\textcolor{black}{
    \begin{aligned}
        G(\mathbf{v}) = \frac{\sqrt{2}G_{\text{F}}}{(2\pi)^3} \int E_\nu^2 dE_\nu \Big[ (f_{\nu_e}(\mathbf{p}) - f_{\nu_x}(\mathbf{p}))
         \\ - (f_{\bar{\nu}_e}(\mathbf{p}) - f_{\bar{\nu}_x}(\mathbf{p})) \Big],
    \end{aligned}}
\end{equation}
crosses zero at some \(\mathbf{v} = \mathbf{v}(\mu, \phi_\nu)\), where \(\mu = \cos\theta_\nu\)~\cite{Morinaga:2021vmc}. Here, \(G_{\rm{F}}\) denotes the Fermi coupling constant, while \(E_\nu\), \(\theta_\nu\), and \(\phi_\nu\) represent the neutrino energy, and the zenith and azimuthal angles of the neutrino velocity, respectively. The \(f_{\nu}\)'s denote neutrino occupation numbers for different flavors, with \(\nu_x\) and \(\bar\nu_x\) referring to heavy-lepton neutrinos and antineutrinos.    \textcolor{black}{Where $\mathbf{p}$ is the momentum of the neutrino.}
Following state-of-the-art CCSNe and NSM simulations, we assume \(\nu_x\) and \(\bar\nu_x\) share similar angular distributions. \textcolor{black}{ Under this assumption, \autoref{eq1} simplifies to the conventional definition of the neutrino electron lepton number (\(\nu\)ELN).}

The extremely short characteristic scales of FFC, in stark contrast to the much coarser resolution of typical hydrodynamic simulations of CCSNe and NSMs, present a substantial challenge for seamless integration. The vast disparity in temporal and spatial scales across the various physical processes further renders direct, first-principles simulations computationally prohibitive. Compounding these difficulties, certain simulation details also require careful attention—for instance, while the equipartition condition appears reasonable for out-going neutrinos, significant deviations arise for incoming neutrinos, affecting the matter profiles \citep{akaho2025comparativetestingsubgridmodels}. A promising strategy is to conduct small-scale FFCs simulations and apply the insights gained to inform larger hydrodynamic models \cite{xiong2020potential,George:2020veu,Li2021a,Just:2022flt,Fernandez:2022yyv,Ehring:2023abs,Ehring:2023lcd}. To address this, extensive research has focused on local dynamical simulations with periodic boundary conditions \cite{Bhattacharyya:2020dhu,Bhattacharyya:2020jpj,Wu:2021uvt,Richers:2021nbx,Zaizen:2021wwl,Richers:2021xtf,Bhattacharyya:2022eed,Grohs:2022fyq,Abbar:2021lmm,Richers:2022bkd,Zaizen:2022cik,Xiong:2023vcm}, where quasistationary flavor states have been observed in the neutrino gas and are well described by analytical models \cite{Xiong:2023vcm}. \textcolor{black}{ Recent studies suggest that when the initial symmetry of the neutrino field is broken, the nature of flavor conversion changes, potentially offering deeper insights into FFCs \cite{Cornelius:2024zsb,George:2024zxz}.} On the theoretical side, a quantum framework describing neutrino interactions with flavor wave quanta has recently been proposed \cite{Fiorillo:2025npi}. On the computational side, an approximate method to predict the asymptotic state of FFCs without solving the full quantum kinetic equations in the three-flavor framework has also been developed \citep{Liu:2025tnf}. These advances offer potential solutions to the computational challenges of fast flavor evolution. 

Nonetheless, fully incorporating FFC effects into realistic CCSNe and NSM simulations remains challenging, as it requires resolving detailed neutrino angular distributions. A practical approach is to approximate neutrino transport using radial moments, thereby reducing the full angular dependence to a few key parameters. The neutrino distribution can then be characterized by radial moments $I_{n,i}$:

\begin{equation}\label{eq2}
I_{n,i} = \frac{E_{\nu,i}^2 \Delta E_{\nu,i}}{(2\pi)^3} \int_{-1}^{1} d\mu\, \mu^n \int_{0}^{2\pi} d\phi_\nu\, f_\nu(p).
\end{equation}

Here, \( E_{\nu,i} \) and \( \Delta E_{\nu,i} \) denote the mean energy and width of the \( i \)-th energy bin, where \( I_{0,i} = n_{\nu,i} \) represents the neutrino count in that bin. Tracking these moments simplifies neutrino transport calculations. In practice, simulations typically provide only $I_{0,i}$ and $I_{1,i}$, but in reality, what we need are their final values determined after FFCs. In addition, the energy-integrated moments can be obtained by summing over all $I_i$ 's.

Although flavor equilibration may occur immediately following rapid FFCs, this is not generally the asymptotic outcome, as feedback from subsequent collisional neutrino-matter interactions becomes increasingly important—particularly when FFCs occur inside the neutrinosphere \citep{PhysRevD.109.123008}. Effective analytical methods have been developed to enable efficient and robust modeling of collisional feedback within the neutrinosphere \citep{PhysRevLett.134.051003}. Nonetheless, leveraging artificial intelligence to uncover hidden physical patterns in data remains a highly promising direction. This is because traditional analytical approaches typically involve specific assumptions and may not fully capture all aspects of the system’s complexity, creating unique opportunities for data-driven methods. With the rapid development of various neutrino telescopes, employing AI techniques offers an effective way to handle the vast volumes of observational data expected in the future and to extract the underlying physical insights. Recently, advances in machine learning, particularly neural networks, have been increasingly applied to astrophysics, including cosmological simulations \citep{2024NatAs...8.1332H,George:2017pmj,CAMELS:2020cof,Nygaard:2024lna} and astronomical data processing \citep{Tripathi:2025vzb,Rezaei:2025nnr,Janulewicz:2025qvp,Fortunato:2024hfm,Fagin:2024qtd,2025arXiv250612230T,George:2016hay,smith2023astronomia,Koundal:2025mxd}. Recent studies have shown that artificial neural networks (NNs) can accurately predict the asymptotic outcomes of FFCs in both single- and multi-energy neutrino gases within the moments framework \citep{Abbar:2023ltx,PhysRevD.109.083019,PhysRevD.109.023033,Richers:2024zit}. 

% While deep learning has achieved remarkable success in certain research areas, many networks are parameter-intensive and lack transparency, making their internal mechanisms difficult to interpret. This remains a key challenge in AI-driven scientific research \citep{nature}. Addressing this issue requires moving beyond traditional \lq \lq black box\rq \rq \ models, as a deeper understanding of phenomena demands interpretability \citep{whitepaper}. 
% Interpretable networks are crucial for processing and analyzing data that may contain FFCs signatures. They not only enhance efficiency but also facilitate the extraction of physical features by tracing the model’s decision rationale and other similar methods, potentially leading to new hypotheses and insights in astronomy and physics.

\textcolor{black}{
While deep learning has achieved remarkable success in certain research areas, many networks are parameter-intensive and lack transparency, making their internal mechanisms difficult to interpret. This remains a key challenge in AI-driven scientific research \citep{nature,rudin2022interpretable}. This issue is often framed as a direct trade-off between accuracy and interpretability \citep{gao2023interpretability}.} \textcolor{black}{
This trade-off is crucial when dealing with data that may contain complex, multivariate, and coupled features such as FFCs. \textcolor{black}{Mitigating} this limitation requires moving beyond \textcolor{black}{conventional} black-box models, because a deeper understanding of the phenomena demands interpretability \citep{hassija2024interpreting}. Interpretable networks are not a compromise for lower accuracy; they are the preferred outcome of a more rigorous model-selection process \citep{majumdar2025accuracy}. Interpretability enhances efficiency and, more importantly, enables the extraction of physical features by making a model’s decision basis traceable. By linking learned patterns to physical principles, interpretable models transform predictive tools into engines of discovery, potentially yielding new hypotheses and insights in astronomy and physics \citep{2025arXiv250323616W,huang2025interpretable}.}

%That view is described by the Rashomon effect \citep{10.5555/3692070.3693812}, which states that for any given problem there typically exists a whole family of distinct models that achieve similarly high accuracy. This reframes the scientist’s core task: the goal is not merely to identify a single most-accurate “black box” \citep{whitepaper}, but to search strategically within this high-performance Rashomon set for models that are also interpretable \citep{semenova2022existence}.

Recently, Kolmogorov-Arnold Networks (KANs) have emerged as a promising solution. Inspired by the Kolmogorov-Arnold representation theorem, KANs introduce learnable activation functions that significantly improve model expressiveness while maintaining interpretability \citep{liu2024kankolmogorovarnoldnetworks}. \textcolor{black}{Compared to conventional neural networks that focus on extracting features from data, KANs offer a more transparent symbolic structure, enabling us to understand the features they learn from the data more directly.} In parallel, the Sparse Identification of Nonlinear Dynamics (SINDy) framework has demonstrated effectiveness in recovering compact, symbolic representations of governing equations from data \citep{2016PNAS..113.3932B}. By promoting sparsity in the candidate function space, SINDy identifies the minimal set of terms required to describe system dynamics, offering a powerful tool for discovering interpretable physical laws from simulation outputs or observations. \textcolor{black}{It should be noted that since our dataset is formulated as a static mapping from initial conditions to asymptotic spectra, we apply SINDy to obtain a sparse functional relation in this reduced representation, and interpret it cautiously as an explicit surrogate rather than a physical time derivative law.}

% \textcolor{pink}{In this work, we aim to develop a data-driven mapping from simulation outputs to underlying physical laws, offering a robust framework for interpreting not only the vast volume of astronomical observations but also the increasingly complex simulation data anticipated in the future. }
\textcolor{black}{In this work, we aim to learn an interpretable, data-driven surrogate that identifies physically meaningful structures implied by simulation outputs, offering a robust framework for interpreting both increasingly complex simulations and astronomical observations in the future.}
Specifically, using simulated FFCs data, we employ KANs and SINDy to reconstruct and predict post-FFC supernova neutrino spectra. Our predictions are based on the first two radial moments of neutrinos angular distributions in each energy bin—quantities typically available in state-of-the-art CCSNe and NSM simulations. 
% \textcolor{pink}{This approach is intended to support the integration of FFCs into advanced CCSNe and NSM models through more interpretable frameworks.} 
\textcolor{black}{This approach represents a step toward more interpretable surrogates for FFC outcomes, with the potential to facilitate their future incorporation into advanced CCSNe and NSM simulations.}

In \autoref{Dataset}, we briefly introduce the structure of the model input. \autoref{Construction of Kolmogorov-Arnold Networks} details the construction of KANs. In \autoref{SINDy}, we describe the working principles of SINDy and how it is applied to the FFCs dataset. The reconstruction performance of KANs and SINDy for post-FFC supernova neutrino spectra is demonstrated in \autoref{Neutrino Energy Spectra Reconstruction and Comparative Analysis}. Subsequently, \autoref{Extracting Physical Information from KAN} presents the extraction of physical insights from both the trained KANs and SINDy models.\textcolor{black}{ In \autoref{Discussion}, we  discuss the trade-off between predictive accuracy and interpretability in KAN and SINDy.} Finally, \autoref{Conclusion} provides a summary of this work.

% \section{Construction of Kolmogorov-Arnold Networks}\label{Construction of Kolmogorov-Arnold Networks}
\section{Model Input}\label{Dataset}

\textcolor{black}{
We first briefly introduce the input dataset used in this work, which is identical to that of \citep{PhysRevD.109.083019} and follows the fast–flavor–conversion (FFC) framework of \citep{Wu:2021uvt}. The data are produced by a 1D-box simulation with a periodic boundary in one spatial direction and translational symmetry in the other two, evolved for several thousand interaction timescales. It spans a grid of initial $\bar\nu_e/\nu_e$ number–density ratios and small seed perturbations, in the two–flavor, forward–scattering limit (neglecting momentum–changing collisions). Although simulations can realize FFCs under simplified conditions, they remain computationally expensive and difficult to embed directly into large-scale CCSN models. A surrogate model offers a physics-aware but lightweight framework that can be integrated into future simulations, while also enabling the extraction of physical insight from observational data or more complex numerical results. Rather than replacing the simulations, the surrogate serves as a bridge between detailed calculations and realistic astrophysical applications.} \textcolor{black}{For training and testing our models, the dataset we used combines maximum entropy and Gaussian initial neutrino angular distributions. The final outcome of FFCs is determined by the two-flavor survival probability:}

\textcolor{black}{
\begin{equation}\label{mupm}
P_{\text{sur}}(\mu) = 
\begin{cases}
\frac{1}{2}, & \text{for } \mu^<  \\
\mathcal{S}_2(\mu), & \text{for } \mu^>
\end{cases},
\end{equation}
where $\mathcal{S}_2(\mu) = 1 - \frac{1}{2} h\left( \left| \mu - \mu_c \right| / \zeta \right)$,$ \quad h(x) = \left(x^2 + 1\right)^{-1/2}$. The parameter \( \zeta \) is chosen to ensure the continuity of \( P_{\text{sur}}(\mu) \). We divide the range of values for $\mu$ into two parts separated by the zero point $\mu_c$, where $\mu_c$ is the position where the $\nu$ELN crossing occurs. The integrals over these two parts are then calculated separately:}

$$
\Gamma_{+} = \left| \int_{-1}^{1} \mathrm{d}\mu\, G(\mu)\, \Theta[G(\mu)] \right|,
$$
$$
\Gamma_{-} = \left| \int_{-1}^{1} \mathrm{d}\mu\, G(\mu)\, \Theta[-G(\mu)] \right|,
$$
\textcolor{black}{where \( \Theta \) denotes the Heaviside theta function, $G(\mu)=\int_0^{2 \pi} \mathrm{~d} \phi_\nu G(\mathbf{v})$ \citep{Xiong:2023vcm,Abbar:2023ltx}.For convenience, we refer to the interval of $\mu$ that makes the above integral small (large) as the "small" ("large") side, which corresponds to $\mu^<$ ($\mu^>$) in \autoref{mupm}}

Each input sample is represented by an 11‐dimensional feature vector, which combines global moment‐integrated quantities and bin‐specific energy information.  Specifically, let:

\begin{multline}\label{input}
\mathbf{x} = \bigl[\,
\alpha_{\mathrm{tot}},\;
F_{\nu_e}^{\mathrm{tot}},\;
F_{\bar\nu_e}^{\mathrm{tot}},\;
n_{\nu_e},\;
F_{\nu_e,i},\;
n_{\bar\nu_e},\;
F_{\bar\nu_e,i},\;\\
n_{\nu_x},\;
F_{\nu_x,i},\;
E_{\nu_e,i},\;
E_{\bar\nu_e,i}
\,\bigr].
\end{multline}

% The first nine components are defined as follows:
% \begin{itemize}
%   \item $\alpha_{\rm tot}$: total ratio of $\bar\nu_e$ to $\nu_e$ number densities;
%   \item $F_{\nu_e}^{\rm tot}$: total flux factor for electron neutrinos;
%   \item $F_{\bar\nu_e}^{\rm tot}$: total flux factor for electron antineutrinos;
%   \item $n_{\nu_e}$: number density of electron neutrinos after integration over all energies;
%   \item $F_{\nu_e,i}$: flux factor of electron neutrinos in the $i$th energy bin;
%   \item $n_{\bar\nu_e}$: number density of electron antineutrinos after integration over all energies;
%   \item $F_{\bar\nu_e,i}$: flux factor of electron antineutrinos in the $i$th energy bin;
%   \item $n_{\nu_x}$: number density of heavy‐lepton neutrinos after integration over all energies;
%   \item $F_{\nu_x,i}$: flux factor of heavy‐lepton neutrinos in the $i$th energy bin.
% \end{itemize}

\begin{table}[htbp]
    \centering
    \caption{Definitions of the physical input variables.}
    \label{tab:var_definitions}
    \renewcommand{\arraystretch}{1.2}
    \setlength{\tabcolsep}{3pt}
    \small

    \begin{tabular}{@{}l l@{}}
        \toprule
        \textbf{Symbol} & \textbf{Description} \\
        \midrule
        $\alpha_{\mathrm{tot}}$ &
        \parbox[t]{0.68\columnwidth}{Total ratio of $I_0^{\bar{\nu}_e}$ to $I_0^{\nu_e}$.} \\

        $F_{\nu_e}^{\mathrm{tot}}$ &
        \parbox[t]{0.68\columnwidth}{Total flux factor for electron neutrinos.} \\

        $F_{\bar{\nu}_e}^{\mathrm{tot}}$ &
        \parbox[t]{0.68\columnwidth}{Total flux factor for electron antineutrinos.} \\

        $n_{\nu_e}$ &
        \parbox[t]{0.68\columnwidth}{Number density of electron neutrinos after integration over all energies.} \\

        $F_{\nu_e,i}$ &
        \parbox[t]{0.68\columnwidth}{Flux factor of electron neutrinos in the $i$th energy bin.} \\

        $n_{\bar{\nu}_e}$ &
        \parbox[t]{0.68\columnwidth}{Number density of electron antineutrinos after integration over all energies.} \\

        $F_{\bar{\nu}_e,i}$ &
        \parbox[t]{0.68\columnwidth}{Flux factor of electron antineutrinos in the $i$th energy bin.} \\

        $n_{\nu_x}$ &
        \parbox[t]{0.68\columnwidth}{Number density of heavy-lepton neutrinos after integration over all energies.} \\

        $F_{\nu_x,i}$ &
        \parbox[t]{0.68\columnwidth}{Flux factor of heavy-lepton neutrinos in the $i$th energy bin.} \\
        \bottomrule
    \end{tabular}
\end{table}

% \textcolor{black}{
% Our 11-dimensional input vector is constructed from two sets of features. 
% The first nine quantities, which primarily describe the moments of the neutrino distribution, are defined in \autoref{tab:var_definitions}. 
% To these, we append two additional energy-dependent components, $E_{\nu_e,i}$ and $E_{\bar\nu_e,i}$, defined for each bin $i$ as follows:"
% }

Our 11-dimensional input vector consists of two sets of features.
The first nine quantities primarily describe the moments of the neutrino distribution, as defined in \autoref{tab:var_definitions}.
\textcolor{black}{To keep the model general, in contrast to \citep{PhysRevD.109.083019} that treats energy bins independently, we include two energy-dependent components,} $E_{\nu_e,i}$ and $E_{\bar\nu_e,i}$, defined for each bin $i$ as follows \citep{PhysRevD.109.083019}:

\begin{equation}\label{energy}
  E_{\nu,i}
  \;=\;
  70 \;-\; 60 \,\sqrt{\frac{F_{\nu,i}}{F_{\nu}^{\rm tot}}}\,,
  \quad
  E_{\nu,i} \in [0,\,60],
\end{equation}
with an identical expression for antineutrinos.  In practice, each $E_{\nu,i}$ is computed from the ratio $F_{\nu,i}/F_{\nu}^{\rm tot}$ and then clipped to the interval $[0,\,60]$ to ensure numerical stability.  A mask is applied to retain only those bins for which both $E_{\nu_e,i}$ and $E_{\bar\nu_e,i}$ are nonnegative; this masking is applied consistently to both inputs and targets so that the original training‐set composition is preserved.  
Finally, $E_{\nu_e,i}$ and $E_{\bar\nu_e,i}$ are concatenated (in that order) to the nine moment‐integrated inputs, yielding an 11‐dimensional vector for each sample.  
\textcolor{black}{
By appending \(E_{\nu_e,i}\) and \(E_{\bar{\nu}_e,i}\) as continuous coordinates of energy, a single unified network can condition its prediction on energy and thereby act as a continuous surrogate across energy, rather than as a per--bin regressor.
Although in the FFCs process the learned dependence on \(E\) is expected to be weak, including these tags is structurally important to maintain generality and portability when subleading, energy--dependent effects become non--negligible (see \autoref{Extracting Physical Information from KAN} for the energy--dependence discussion).
}

% Throughout this process, the input dataset remains unchanged except for the bin‐mask filtering, thereby ensuring that the models learn from a physically consistent input distribution.

The model parameters are as follows:  
\( \alpha \in (0, 2.5) \),  
\( F_{\nu_x, (i)} \in (0, 1) \),  
\( \bar{F}_{\nu_e, (i)} \in (0.4 F_{\nu_x, (i)}, F_{\nu_x, (i)}) \), and  
\( F_{\nu_e, (i)} \in (0.4 \bar{F}_{\nu_e, (i)}, \bar{F}_{\nu_e, (i)}) \),  
which follows the expected hierarchy:  
\( F_{\nu_e} \lesssim \bar{F}_{\nu_e} \lesssim F_{\nu_x} \). The neutrino number densities \( n_{\nu, i} \) are sampled from a half-normal distribution with a mean of zero and a standard deviation of \( 0.1n_\nu \), where \( n_\nu \) is the energy-integrated neutrino number density. \textcolor{black}{ With \( F_{\nu} = (I_1/I_0)_{\nu} \) and $\alpha = n_{\bar{\nu}_{e}}/n_{\nu_{e}}$,  the quantities with and without the subscript \( i \) refer to the energy-integrated spectrum and specific energy bin, respectively.} The neutrino number densities in a given energy bin must be smaller than the corresponding energy-integrated values. This is consistent with \citep{PhysRevD.109.083019}.
\textcolor{black}{Moreover, the dataset used for both modeling approaches is consistent with \citep{PhysRevD.109.083019}, including the sample size, the train/validation split for the KAN network in \autoref{Construction of Kolmogorov-Arnold Networks}, and other detailed settings. In particular, the SINDy method in \autoref{SINDy} does not require a train/validation split and is fitted directly on the full dataset. Because the core of SINDy is to use sparse regression to select as few non-zero coefficients as possible from the candidate library, resulting in the "simplest" control equations; this step itself is a structural selection and does not depend on training/validation splitting.}

\section{Kolmogorov-Arnold Networks}\label{Construction of Kolmogorov-Arnold Networks}

Kolmogorov-Arnold representation theorem states that if $f$ is a multivariate continuous function on a bounded domain, then $f$ can be written as a finite composition of continuous functions of a single variable and the binary operation of addition. More specifically, for a smooth $f:[0, 1]^n \rightarrow \mathbb{R}$:

\begin{equation}\color{black}
f(x_1, \ldots, x_n) = \sum_{q=1}^{2n+1} \Phi_q \left( \sum_{p=1}^n \phi_{q,p}(x_p) \right)
\label{kanRT}.
\end{equation}

\textcolor{black}{Similar to Multilayer Perceptrons (MLPs), which are based on the Universal Approximation Theorem, KANs integrate the Kolmogorov-Arnold representation theorem into neural networks. Instead of relying on fixed activation functions between nodes, KANs employ activation functions parameterized by trainable B-splines, making them learnable during training. As a result, both \( \Phi_q \) and \( \phi_{q,p} \) in \autoref{kanRT} become functions that the network can adaptively learn. Specifically, at each layer, the transformation \( \Phi_l \) operates on the input \( x_l \) to produce the input for the next layer, \( x_{l+1} \), as represented in the following matrix:}

\begin{multline}\color{black}
    {x}_{l+1} = \Phi_l(x_l) \\ =
    \begin{pmatrix}
        \phi_{l,1,1}(\cdot) & \phi_{l,1,2}(\cdot) & \cdots & \phi_{l,1,n_{l}}(\cdot) \\
        \phi_{l,2,1}(\cdot) & \phi_{l,2,2}(\cdot) & \cdots & \phi_{l,2,n_{l}}(\cdot) \\
        \vdots & \vdots & & \vdots \\
        \phi_{l,n_{l+1},1}(\cdot) & \phi_{l,n_{l+1},2}(\cdot) & \cdots & \phi_{l,n_{l+1},n_{l}}(\cdot) \\
    \end{pmatrix}\label{eq:kanforwardmatrix}
    {x}_{l}.
\end{multline}

\textcolor{black}{The final output of the network is given by the composition of all layer transformations, as shown in the following equation:}
\begin{equation}\color{black}
    \text{KANs}(x) = (\Phi_{L-1} \circ \Phi_{L-2} \circ \cdots \circ \Phi_0)(x).
\end{equation}

\textcolor{black}{We can also rewrite the above equation to make it more analogous to \autoref{kanRT}, assuming output dimension $n_{L}=1$, and define $f({x})\equiv {\rm KAN}({x})$:}

% \begin{equation}\color{black}
%     f({x})=\sum_{i_{L-1}=1}^{n_{L-1}}\phi_{L-1,i_{L},i_{L-1}}\left(\sum_{i_{L-2}=1}^{n_{L-2}}\cdots\left(\sum_{i_2=1}^{n_2}\phi_{2,i_3,i_2}\left(\sum_{i_1=1}^{n_1}\phi_{1,i_2,i_1}\left(\sum_{i_0=1}^{n_0}\phi_{0,i_1,i_0}(x_{i_0})\right)\right)\right)\cdots\right)
% \end{equation}

\begin{multline}
    f(x) = \sum_{i_{L-1}=1}^{n_{L-1}} \phi_{L-1,i_{L},i_{L-1}} \Bigg( \cdots \\
    \left( \sum_{i_1=1}^{n_1} \phi_{1,i_2,i_1} \left( \sum_{i_0=1}^{n_0} \phi_{0,i_1,i_0}(x_{i_0}) \right) \right) \cdots \Bigg).
    \label{eq:kan_simplified}
\end{multline}

\textcolor{black}{Compared to MLPs with fixed activation functions, these diverse set of \( \{\phi_{l,j,k}\} \) not only enables the extraction of richer nonlinear features from the data through training but can also be further fitted to elementary functions, which enables KAN-based networks to achieve the same level of accuracy with fewer nodes and parameters, thereby enhancing interpretability \citep{liu2024kankolmogorovarnoldnetworks}.}

% \section{Network Architecture and Training Configuration}

\begin{figure}[H]
    \centering % 将图片在栏内居中
    \includegraphics[width=1\columnwidth]{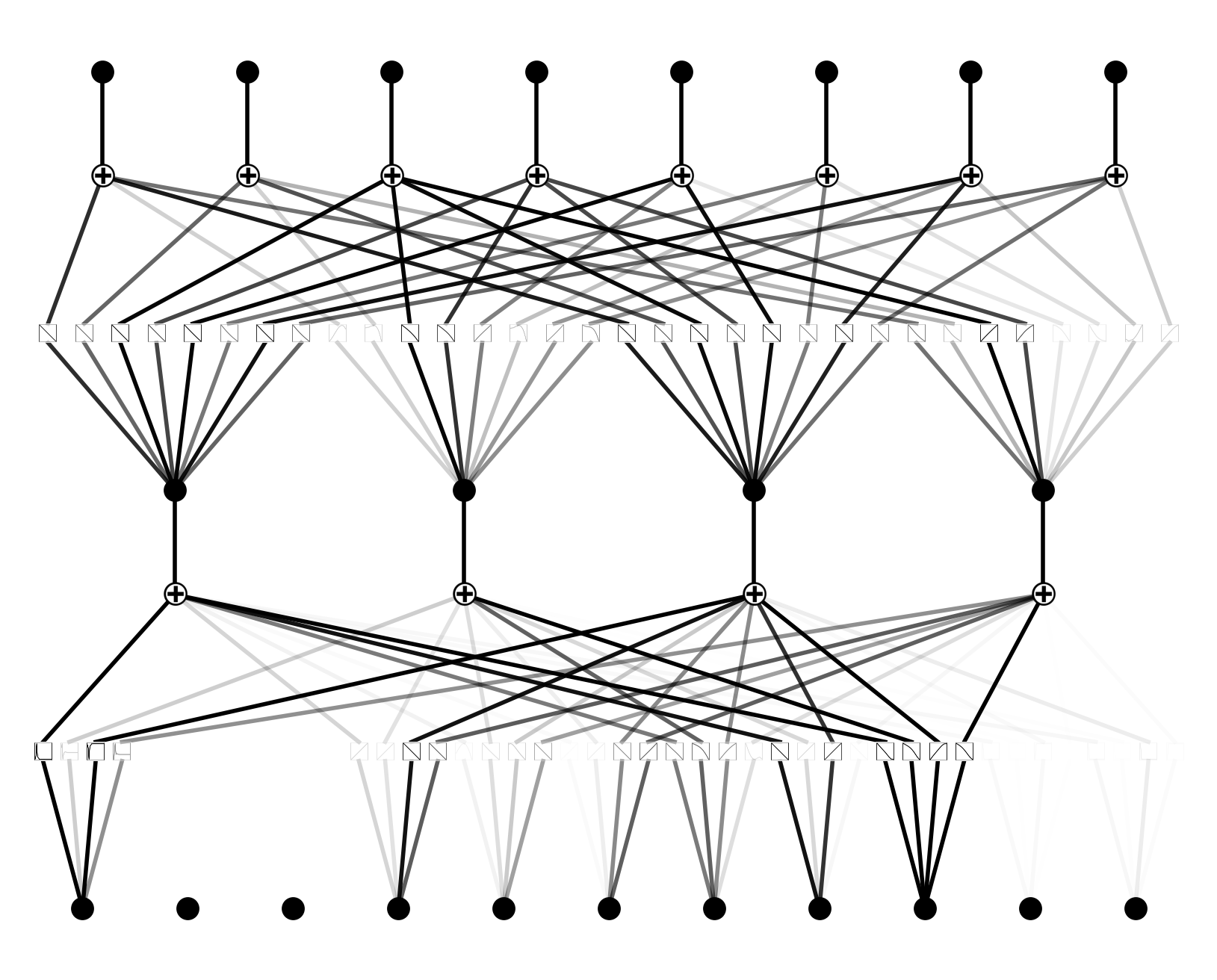}
    \caption{
    Visualization of the trained KAN architecture used to model the fast flavor conversion process.  
    The eleven input neurons at the bottom correspond, from left to right, to the physical features of the pre-conversion neutrino gas as detailed in \autoref{input}. 
    The eight output neurons at the top yield the predicted post-conversion radial moments for all relevant flavors ($I_{0}^{\nu_e}, I_{1}^{\nu_e}, I_{0}^{\bar{\nu}_e}, I_{1}^{\bar{\nu}_e}, I_{0}^{\nu_x}, I_{1}^{\nu_x}, I_{0}^{\bar{\nu}_x}, I_{1}^{\bar{\nu}_x}$, from left to right). 
    The curves on the edges represent the learned one-dimensional activation functions, which are parameterized by B-splines. 
    The activation shapes shown are taken from a representative training run, illustrating the complex functional relationships captured by the model.
    \textcolor{black}{
    Edge darkness denotes the contribution of the connected input (darker means higher contribution); nearly invisible edges indicate negligible contribution.
    }
    }
    \label{kan}
\end{figure}

Tailored to the input dataset detailed in \autoref{Dataset}, the network we designed is shown in \autoref{kan}, comprising 11 input neurons, 4 hidden neurons, and 8 output neurons. Notably, we adopted a unified modeling strategy. 
Our single network architecture simultaneously processes all neutrino channels without decoupling flavors (e.g., electron-neutrinos from their antiparticles) and is applied universally across all energy bins. 
This holistic approach ensures that the network, which will later undergo symbolic regression, can more effectively capture the underlying correlations and dynamics of the FFC process.

In our experiments, the network is trained with a composite loss function that balances mean squared error against three physics‐inspired constraints.  Let $\hat I_{c}$ and $I_{c}$ denote the predicted and true moments for channel $c\in\{1,\dots,8\}$.  We defined the base loss by:

\begin{equation}
L_{\mathrm{base}} \;=\; \sum_{c=1}^8 \bigl(\hat I_{c} - I_{c}\bigr)^2 \,.
\end{equation}

To impose energy–flavor correlations arising from fast flavor conversion, we introduce two combinations:

\begin{equation}
a_{\nu_e e} \;=\; E_{\nu_e}\,I_{0}^{\nu_e} \;+\; E_{\bar\nu_e}\,I_{0}^{\bar\nu_e},
\end{equation}
\begin{equation}
a_{\nu_e x} \;=\; E_{\nu_e}\,I_{0}^{\nu_x} \;-\; E_{\bar\nu_e}\,I_{0}^{\bar\nu_x},
\end{equation}
where $E_{\nu_e}$ and $E_{\bar\nu_e}$ are the energy features introduced in \autoref{energy}.  We then define two pointwise auxiliary losses:

\begin{equation}
L_{\mathrm{extra},1} \;=\; \bigl(a_{\nu_e e}^{\mathrm{pred}} - a_{\nu_e e}^{\mathrm{true}}\bigr)^2,
\end{equation}
\begin{equation}
L_{\mathrm{extra},2} \;=\; \bigl(a_{\nu_e x}^{\mathrm{pred}} - a_{\nu_e x}^{\mathrm{true}}\bigr)^2,
\end{equation}
together with a global‐exchange term that matches the batch‐mean of $a_{\nu_e x}$ between prediction and target:

\begin{equation}
L_{\mathrm{extra},3} \;=\; \Bigl(\,\overline{a_{\nu_e x}^{\mathrm{pred}}} \;-\; \overline{a_{\nu_e x}^{\mathrm{true}}}\Bigr)^2.
\end{equation}

Rather than fixing the relative weights of $L_{\mathrm{extra},1}$, $L_{\mathrm{extra},2}$, and $L_{\mathrm{extra},3}$ by hand, we employ a small ancillary module, LossMixer, whose sole trainable parameters are a vector $\bm\alpha\in\mathbb{R}^3$.  At each training step, the softmax of $\bm\alpha$ yields a probability vector $\bm p=(p_1,p_2,p_3)$ that adaptively weights the three auxiliary losses:

\begin{equation}
L_{\mathrm{aux}} \;=\; p_1\,L_{\mathrm{extra},1} \;+\; p_2\,L_{\mathrm{extra},2} \;+\; p_3\,L_{\mathrm{extra},3}.
\end{equation}

Finally, the total loss is defined as:

\begin{equation}
L_{\mathrm{total}} \;=\; 0.9\,L_{\mathrm{base}} \;+\; 0.1\,L_{\mathrm{aux}}.
\end{equation}

This formulation ensures that the network learns to reproduce all eight output moments accurately (via $L_{\mathrm{base}}$) while simultaneously enforcing the energy–flavor conservation relationships (via the three auxiliary terms).  Empirically, the dynamic reweighting implemented by LossMixer accelerates convergence, yields predictions that honor underlying physical constraints, and improves generalization compared to using an unweighted mean squared error alone.

% \begin{figure}[H]
%     \centering
%     \includegraphics[width=1\linewidth]{metrics_plot.png}
%     \caption{Validation metrics of the network. The plot shows the evolution of relative error on the validation set as a function of the training epoch, with the y-axis on a logarithmic scale. The blue line (circles) represents the overall relative error of the network, while the orange line (squares) represents the specific error component associated with the electron neutrino and antineutrino terms ($\nu_e + \bar{\nu}_e$).}
%     \label{metrics}
% \end{figure}

\begin{figure}[H]
    \centering % 将图片在栏内居中
    \includegraphics[width=\columnwidth]{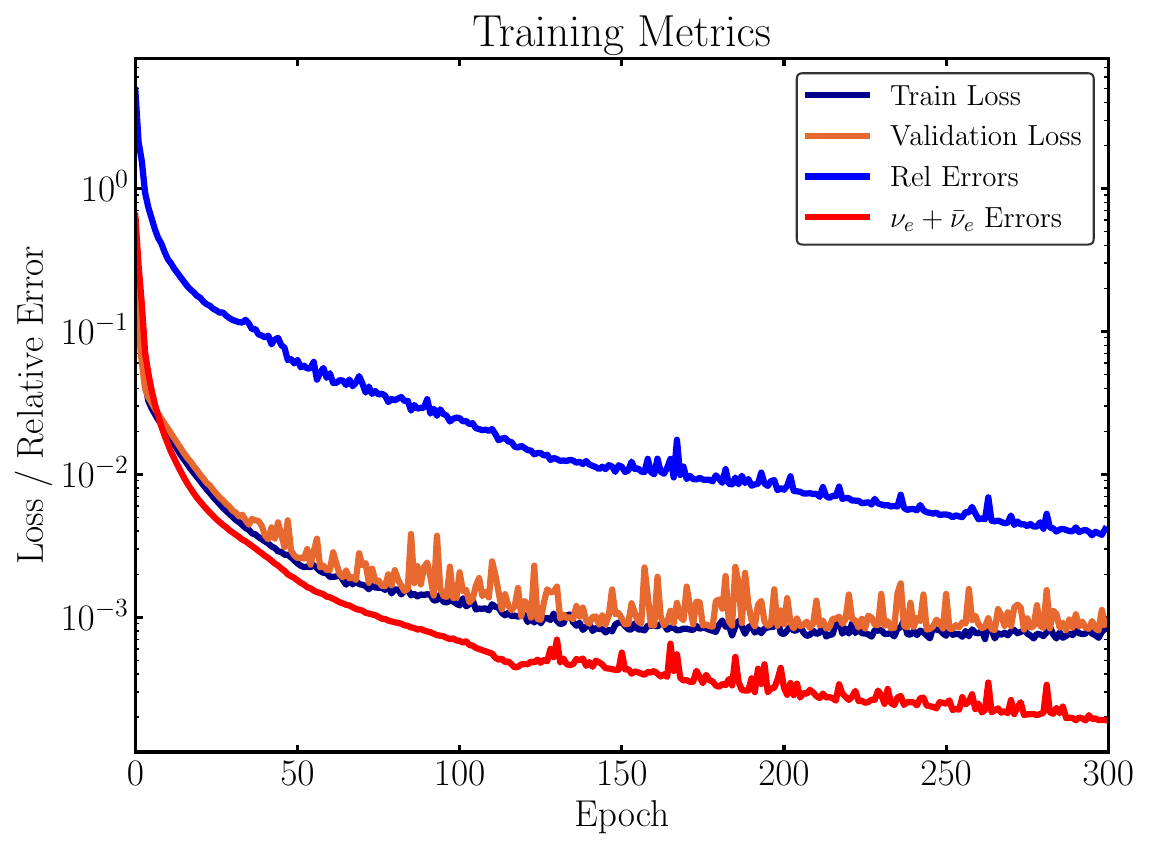}
    \caption{
        The evolution of key metrics during the training process. 
        The plot displays the training loss, validation loss, and two physics-informed error terms as a function of the training epoch. 
        The logarithmic scale on the y-axis highlights the convergence behavior of the network.
    }
    \label{fig:training_metrics}
\end{figure}

{\color{black}
We also define two terms used to assess model performance. Both metrics are forms of the Mean Absolute Error (MAE) computed on the validation set. A small constant $\epsilon=10^{-8}$ is added to the denominator to ensure numerical stability.

The overall relative error, denoted $\mathcal{L}_{\text{rel}}$, measures the average relative error across all eight output neurons. For a validation set with $N$ samples, where $\hat{y}_{j,i}$ is the $i$-th predicted output for sample $j$ and $y_{j,i}$ is the corresponding true value,
\begin{equation}
    \mathcal{L}_{\text{rel}} = \frac{1}{N}\sum_{j=1}^{N}\left(\frac{1}{8}\sum_{i=0}^{7}\frac{|\hat{y}_{j,i}-y_{j,i}|}{|y_{j,i}|+\epsilon}\right).
    \label{eq:L_rel}
\end{equation}

The $\boldsymbol{\nu_e + \bar{\nu}_e}$ error, denoted $\mathcal{L}_{\nu}$, evaluates the model’s ability to predict the total contribution of electron neutrinos and antineutrinos. For each sample $j$, let $S_{\text{pred},j}$ and $S_{\text{true},j}$ be the predicted and true totals, respectively. The loss is
\begin{equation}
    \mathcal{L}_{\nu} = \frac{1}{N}\sum_{j=1}^{N}\frac{|S_{\text{pred},j}-S_{\text{true},j}|}{|S_{\text{true},j}|+\epsilon}.
    \label{eq:L_nu}
\end{equation}

\autoref{fig:training_metrics} then reports these metrics over the course of training: the blue curve corresponds to $\mathcal{L}_{\text{rel}}$ and the red curve to $\mathcal{L}_{\nu}$.
}

% --- 创建一个单栏的浮动体环境来插入训练曲线图 ---

\autoref{kan} visualizes the architecture of the trained KAN from a representative run. The opacity of each edge connecting the nodes is proportional to the amplitude of the corresponding B-spline activation function learned on that path. This amplitude can be understood as representing the relative contribution of the corresponding node within the entire model. A key feature is the network's intrinsic pruning of several connections, which are visualized as faded or white lines. Notably, connections originating from the second, third, tenth, and eleventh input neurons are systematically discarded. These inputs correspond to the total flux factors ($F_{\nu_e}^{\mathrm{tot}}$, $F_{\bar{\nu}_e}^{\mathrm{tot}}$) and the bin-specific energies ($E_{\nu_e,i}$, $E_{\bar{\nu}_e,i}$), respectively. We found this result to be a robust feature, consistently emerging across multiple training initializations with different random seeds, rather than an artifact of a single run. This interesting finding is discussed further in \autoref{Extracting Physical Information from KAN}.

% However, during the symbolic fitting of KAN neurons, we observed that the choice of random seed exerts a non‐negligible influence on the fitting outcomes. To mitigate this randomness‐induced bias, we conducted multiple training runs and computed the mean value of each B-spline parameters across those runs. We then used these averaged parameters to perform symbolic regression for each neuron. Throughout this procedure, the training dataset remained unchanged.

% However, we identified that the symbolic regression process, a key step for KAN interpretability, exhibits a notable sensitivity to the random seed used in training. To obtain a robust and canonical symbolic model, we adopted an ensemble averaging approach. Specifically, the network was trained multiple times with different random seeds, and the B-spline coefficients for each learnable activation function were averaged across all runs. By performing symbolic regression on this stable set of averaged parameters, we mitigate the effects of stochastic initialization and derive a more representative model of the underlying dynamics.
{\color{black}
In the KAN framework, each learnable activation is parameterized by B-splines, which can subsequently be approximated by elementary analytic functions through the built-in symbolic regression module. 
This procedure enhances interpretability by providing symbolic results of the learned nonlinearities without requiring any modification of the original KAN implementation.
In our study, we used the default symbolic regression functionality included in the KAN library \citep{liu2024kankolmogorovarnoldnetworks}. 

However, the symbolic regression process exhibits a notable sensitivity to the random seed used in training. To obtain a robust and canonical symbolic model, we adopted an ensemble averaging approach. Specifically, the network was trained multiple times with different random seeds, and the B-spline coefficients for each learnable activation function were averaged across all runs. By performing symbolic regression on this stable set of averaged parameters, we mitigate the effects of stochastic initialization and derive a more representative model of the underlying dynamics.

% The resulting analytic expressions are provided in Appendix~\ref{app:kan_sr} for reference. Compared with the PySINDy results presented in Section~IV, the expressions extracted from KANs contain considerably more terms, which is why we describe them as too complicated for practical physical interpretation.
}

\textcolor{black}{\section{Sparse Identification of Nonlinear system part}}\label{SINDy}

% In this study, we employ the Sparse Identification of Nonlinear Dynamics (SINDy) framework to extract interpretable models from time-series data. SINDy assumes that the system dynamics can be described by a sparse combination of candidate nonlinear functions.

% Although KANs can successfully build analytical network for the neutrino fast flavor conversion, we identified two key limitations: the resulting formulas can lack the desired simplicity, and the process of symbolic regression sometimes struggles with randomness-induced bias. 

% \textcolor{pink}{Although KANs can successfully build analytical network for the neutrino fast flavor conversion} 
\textcolor{black}{Although KANs provide a structured functional surrogate for fast flavor conversion outcomes,} we identified two key limitations: the resulting formulas \textcolor{black}{remain relatively complex}, and the process of symbolic regression sometimes struggles with randomness-induced bias.
To overcome these specific challenges and to construct a more concise, low-rank model, we employ the Sparse Identification of Nonlinear Dynamics (SINDy) framework to extract interpretable models from input data. 
SINDy assumes that the system dynamics can be described by a sparse combination of candidate nonlinear functions.
The system is modeled as a general nonlinear dynamical system \citep{2016PNAS..113.3932B}:
\begin{equation}
\dot{\mathbf{x}}(t) = \mathbf{f}(\mathbf{x}(t)),
\label{eq:sindy_dynamics}
\end{equation}
where $\mathbf{x}(t) \in \mathbb{R}^n$ is the state vector and $\mathbf{f}$ is an unknown function representing the dynamics.

To identify $\mathbf{f}$ from data, we collect a sequence of measurements $\mathbf{x}(t_1), \dots, \mathbf{x}(t_m)$ and their derivatives $\dot{\mathbf{x}}(t_1), \dots, \dot{\mathbf{x}}(t_m)$, and form the data matrices:
% \begin{equation}
% \mathbf{X} = 
% \begin{bmatrix}
% \mathbf{x}(t_1)^T \\
% \vdots \\
% \mathbf{x}(t_m)^T
% \end{bmatrix}, \quad
% \dot{\mathbf{X}} = 
% \begin{bmatrix}
% \dot{\mathbf{x}}(t_1)^T \\
% \vdots \\
% \dot{\mathbf{x}}(t_m)^T
% \end{bmatrix}.
% \end{equation}

\begin{equation}\label{sindy_ori1}
X = 
\begin{bmatrix}
\mathbf{x}^T(t_1) \\
\mathbf{x}^T(t_2) \\
\vdots \\
\mathbf{x}^T(t_m)
\end{bmatrix} =
\begin{bmatrix}
x_1(t_1) & x_2(t_1) & \cdots & x_n(t_1) \\
x_1(t_2) & x_2(t_2) & \cdots & x_n(t_2) \\
\vdots & \vdots & \ddots & \vdots \\
x_1(t_m) & x_2(t_m) & \cdots & x_n(t_m)
\end{bmatrix},
\end{equation}

\begin{equation}\label{sindy_ori2}
\dot{\mathbf{X}} = 
\begin{bmatrix}
\dot{\mathbf{x}}^T(t_1) \\
\dot{\mathbf{x}}^T(t_2) \\
\vdots \\
\dot{\mathbf{x}}^T(t_m)
\end{bmatrix} =
\begin{bmatrix}
\dot{x}_1(t_1) & \dot{x}_2(t_1) & \cdots & \dot{x}_n(t_1) \\
\dot{x}_1(t_2) & \dot{x}_2(t_2) & \cdots & \dot{x}_n(t_2) \\
\vdots & \vdots & \ddots & \vdots \\
\dot{x}_1(t_m) & \dot{x}_2(t_m) & \cdots & \dot{x}_n(t_m)
\end{bmatrix} .
\end{equation}

Since our dataset corresponds to a neutrino system that has undergone FFCs and settled into the asymptotic state, \autoref{sindy_ori1} and \autoref{sindy_ori2} reduce to:

\begin{equation}\label{sindy_ori3}
\mathbf{X} = 
\begin{bmatrix}
\mathbf{x}_{(1)}^{T} \\
\vdots \\
\mathbf{x}_{(m)}^{T}
\end{bmatrix}, \quad
\dot{\mathbf{X}} =
\begin{bmatrix}
\dot{\mathbf{x}}_{(1)}^{T} \\
\vdots \\
\dot{\mathbf{x}}_{(m)}^{T}
\end{bmatrix}.
\end{equation}

A library of candidate nonlinear functions, $\boldsymbol{\Theta}(\mathbf{X})$, is then constructed by evaluating various functions on $\mathbf{X}$:
\begin{equation}
\boldsymbol{\Theta}(\mathbf{X}) = 
\begin{bmatrix}
\mathbf{1} & \mathbf{X} & \mathbf{X}^{P_2} & \cdots & \sin(\mathbf{X}) & \cos(\mathbf{X}) & \cdots
\end{bmatrix},
\label{eq:sindy_library}
\end{equation}
where $\mathbf{1}$ represents a column of ones for constant terms, $\mathbf{X}^{P_2}$ represents second-order polynomial terms (e.g., $x_i^2$, $x_ix_j$), and other functions like trigonometric terms are included as deemed relevant for the specific dynamics.

The dynamics are then approximated by a sparse linear combination of these candidate functions:
\begin{equation}
\dot{\mathbf{X}} = \boldsymbol{\Theta}(\mathbf{X}) \boldsymbol{\Xi},
\label{eq:sindy_regression}
\end{equation}
where $\boldsymbol{\Xi}$ is a sparse coefficient matrix. Each column of $\boldsymbol{\Xi}$ identifies the active terms in the corresponding component of the dynamics. 

A critical challenge within the traditional SINDy method is the selection of a suitable sparsification threshold. To obtain a sparse coefficient matrix $\boldsymbol{\Xi}$ in \autoref{eq:sindy_regression}, a threshold is typically applied to eliminate terms with coefficient values below a certain magnitude. However, the choice of this threshold is decisive for the correct identification of the model structure, and an inappropriate threshold can lead to inaccurate model identification or significantly increased computational costs.

To circumvent this limitation, we use Global Sensitivity Analysis (GSA) techniques, forming the SINDy-SA framework \citep{naozuka2022sindy}. The core idea of SINDy-SA is to leverage sensitivity analysis to hierarchize the importance of terms within the candidate function library $\boldsymbol{\Theta}(\mathbf{X})$, thereby circumventing the need for manually defining a sparse threshold.

The SINDy-SA framework operates as an iterative process where sensitivity analysis guides the sparsification. This framework incorporates a general workflow encompassing different experimental settings, recalibration of each identified model, and the use of model selection techniques to choose the optimal and most parsimonious model. Specifically, this iterative workflow encompasses the following key steps:

The process begins with ridge regression estimation. At the start of each iteration, an estimate for the coefficient matrix $\boldsymbol{\Xi}$ is obtained by solving a least-squares problem with $l_2$-regularization (ridge regression). Ridge regression is chosen for its robustness and accuracy in balancing the residual sum of squares and preventing overfitting. The estimate for each column $k$ of $\boldsymbol{\Xi}$, denoted $\hat{\boldsymbol{\xi}}_k$, is given by:
\begin{multline}
\hat{\boldsymbol{\xi}}_{k} = \arg\min_{\boldsymbol{\xi}'_k} ||\dot{\mathbf{X}}_k - \boldsymbol{\Theta}(\mathbf{X})\boldsymbol{\xi}'_k||_2 + \alpha||\boldsymbol{\xi}'_k||_2 \text{,} \\ \text{for each column } k \text{ of } \dot{\mathbf{X}} \text{,}
\end{multline}
where $\alpha$ is the tuning parameter that controls the relative impact of the two terms on the coefficient estimates.

Following ridge regression, error computation and convergence check are performed. The predicted derivatives $\hat{\dot{\mathbf{X}}} = \boldsymbol{\Theta}(\mathbf{X})\hat{\boldsymbol{\Xi}}$ are computed, and the Sum of Squared Errors (SSE) between the measured or approximated derivatives $\dot{\mathbf{X}}$ and the predicted derivatives $\hat{\dot{\mathbf{X}}}$ is calculated. The iterative process continues until a significant increase in the SSE is observed compared to the mean ($\mathcal{M}_\tau$) and standard deviation ($\mathcal{D}_\tau$) of errors from previous iterations. This stopping criterion ensures that further elimination of terms does not unduly compromise model accuracy.

If the SSE does not significantly increase, the Sensitivity Analysis is performed on the current model parameters (i.e., the non-zero coefficients in $\boldsymbol{\Xi}$). This study specifically utilizes the Morris method, also known as the Elementary Effects (EE) method, due to its simplicity and low computational cost. The EE method is capable of assessing both the overall importance of the coefficients and their nonlinear and interaction effects. For each input parameter $i$ and each state variable (Quantity of Interest, QoI) $k$, we compute the sensitivity indices $\mu^*_{k,i}$ (representing influence) and $\sigma_{k,i}$ (representing non-linearity/interaction). To integrate these two aspects and facilitate the ranking of parameters by importance, a combined sensitivity index $\mathcal{S}_{k,i}$ is defined as:
$$\mathcal{S}_{k,i} = \sqrt{(\mu^*_{k,i})^2 + (\sigma_{k,i})^2} \text{}.$$

A larger value of $\mathcal{S}_{k,i}$ indicates a greater overall influence of the $i$-th parameter on the $k$-th QoI. Based on these sensitivity indices, candidate functions for each dynamical system equation are ordered, and the least important terms are iteratively eliminated. This procedure ensures that influential terms are retained while achieving model sparsification.

Through the SINDy-SA framework, the model adaptively selects the most relevant terms based on data and the physical importance of the terms, thereby enabling more robust, accurate, and interpretable identification of nonlinear dynamical systems without relying on arbitrary manual thresholds. The specific implementation in this study uses the Python packages \texttt{pySINDy} \citep{desilva2020,Kaptanoglu2022} and \texttt{SALib} \citep{Herman2017,Iwanaga2022}.

\textcolor{black}{Because our dataset contains only the correspondence between initial conditions and asymptotic states (rather than a full time series), the SINDy notation ($\dot{\mathbf{x}}$) is used purely as a formal placeholder in this work: in practice, we replace it with the output vector ($\mathbf{y}$), and thus learn a sparse, closed-form approximation to the mapping from input ($\mathbf{x}$) to output ($\mathbf{y}$). We emphasize that this expression should be interpreted as an explicit surrogate for the asymptotic mapping, rather than as the true time-evolution equations of the underlying neutrino system.} The SINDy and SINDy-SA analysis revealed a concise linear relationship governing the system, which connects the 11 inputs to the 8 outputs. This relationship is captured by the matrix equation:

% \begin{equation}
% \left[\begin{array}{c}
% y_1 \\
% y_2 \\
% \vdots \\
% y_8
% \end{array}\right]=\left[\begin{array}{c}
% c_1 \\
% c_2 \\
% \vdots \\
% c_8
% \end{array}\right]+\left[\begin{array}{cccc}
% A_{1,1} & A_{1,2} & \cdots & A_{1,11} \\
% A_{2,1} & A_{2,2} & \cdots & A_{2,11} \\
% \vdots & \vdots & \ddots & \vdots \\
% A_{8,1} & A_{8,2} & \cdots & A_{8,11}
% \end{array}\right]\left[\begin{array}{c}
% \sin \left(x_1\right) \\
% \sin \left(x_2\right) \\
% \vdots \\
% \sin \left(x_{11}\right)
% \end{array}\right]
% \end{equation}

\begin{equation}
\left[\begin{array}{c}
y_1 \\
y_2 \\
\vdots \\
y_8
\end{array}\right]=\mathbf{A}\left[\begin{array}{c}
x_1 \\
x_2 \\
\vdots \\
x_{11}
\end{array}\right]+\mathbf{B}\left[\begin{array}{c}
\sin \left(x_1\right) \\
\sin \left(x_2\right) \\
\vdots \\
\sin \left(x_{11}\right)
\end{array}\right]+\mathbf{C}.
\end{equation}

This expression defines a model that maps an input vector $\mathbf{x} \in \mathbb{R}^{11}$ to an output vector $\mathbf{y} \in \mathbb{R}^8$. The model consists of a linear mapping of the raw inputs by a coefficient matrix $\mathbf{A} \in \mathbb{R}^{8 \times 11}$, a non-linear component from a matrix $\mathbf{B} \in \mathbb{R}^{8 \times 11}$ that acts on the element-wise sine of the inputs, $\sin (\mathbf{x})$, and an added bias vector $\mathbf{C} \in$ $\mathbb{R}^8$. Here, $\mathbf{x}$ and $\mathbf{y}$ are consistent with the inputs and outputs of the KANs network. The full expressions and coefficients can be found in the link provided in \textit{Code and Data Availability}.

\section{Neutrino Energy Spectra Reconstruction and Comparative Analysis}
\label{Neutrino Energy Spectra Reconstruction and Comparative Analysis}

% \textcolor{black}{In the preceding sections, we have established two distinct data-driven frameworks for modeling fast flavor conversions.
% The ultimate test of these models lies in their ability to perform the crucial task of reconstructing the final, post-conversion neutrino energy spectra from a limited set of initial radial moments. 
% In this section, we apply both the KAN and SINDy models to this reconstruction task. 
% We will first outline the procedure for constructing the energy spectra from the model outputs, and then present a detailed comparative analysis of the models' accuracy and interpretability, using the results to highlight the fundamental trade-offs between these competing objectives.
% }

In the preceding sections, we established two distinct data-driven frameworks for modeling fast flavor conversions. The ultimate test of these models lies in their ability to perform the crucial task of reconstructing the final, post-conversion neutrino energy spectra from a limited set of initial radial moments in the dataset. In this section, we apply both the KAN and SINDy models to this reconstruction task. We first outline the procedure for generating the energy spectra from the model outputs, and then present a detailed comparative analysis of each model’s accuracy and interpretability, using the results to highlight the fundamental trade-offs between these competing objectives.

The energy-differential number flux of a neutrino species \(\nu_\beta\) is expressed as \citep{Mirizzi:2015eza}
\begin{equation}
    \mathcal{F}_{\nu_\beta}(E_\nu) \propto \frac{L_{\nu_\beta}}{\langle E_{\nu_\beta} \rangle} f_{\nu_\beta}(E_\nu),
\end{equation}
where the normalized energy spectrum \(f_{\nu_\beta}(E_\nu)\) is given by
\begin{equation}
    f_{\nu_\beta}(E_\nu) = \frac{1}{T_{\nu_\beta} \, \Gamma(1 + \eta_{\nu_\beta})} 
    \left( \frac{E_\nu}{T_{\nu_\beta}} \right)^{\eta_{\nu_\beta}} 
    \exp\left( - \frac{E_\nu}{T_{\nu_\beta}} \right),
\end{equation}
with \(\eta_{\nu_\beta}\) denoting the spectral pinching parameter, and \(T_{\nu_\beta} = \langle E_{\nu_\beta} \rangle / (1 + \eta_{\nu_\beta})\). Here, \(\langle E_{\nu_\beta} \rangle\) is the mean neutrino energy, and \(L_{\nu_\beta}\) is the total luminosity.

For the SN accretion phase, the adopted benchmark parameters are \citep{Mirizzi:2015eza}:
\begin{equation}
\begin{aligned}
    L_{\nu_e} : L_{\bar{\nu}_e} : L_{\nu_x} &= 1 : 1 : 0.33, \\
    \langle E_{\nu_e} \rangle : \langle E_{\bar{\nu}_e} \rangle : \langle E_{\nu_x} \rangle &= 9 : 12 : 16.5, \\
    \eta_{\nu_e} : \eta_{\bar{\nu}_e} : \eta_{\nu_x} &= 3.2 : 4.5 : 2.3.
\end{aligned}
\end{equation}

\begin{figure}[H]
    \centering
    \includegraphics[width=1\linewidth]{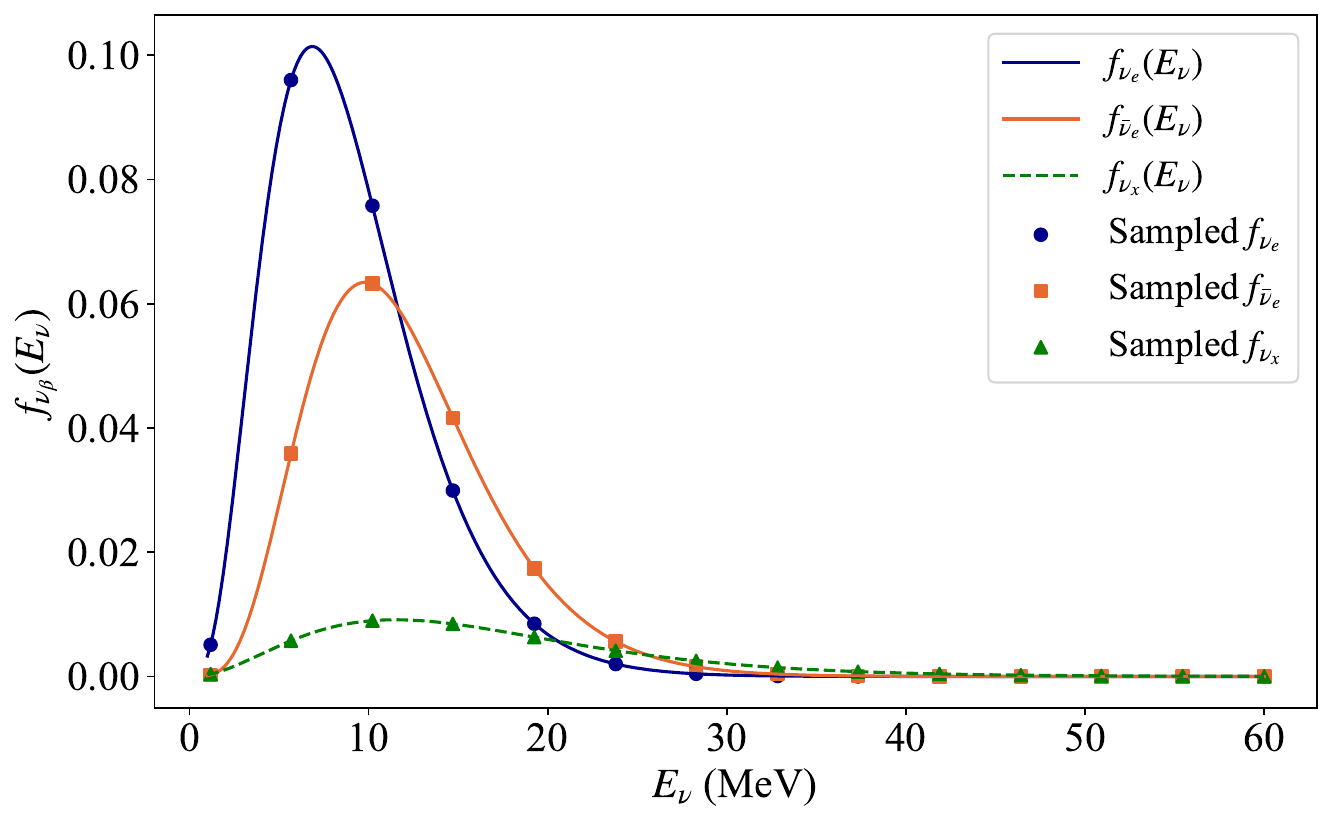}
    \caption{The plot illustrates the energy distributions of three types of neutrinos: electron-type neutrinos (\( \nu_e \)), anti-electron neutrinos (\( \bar{\nu}_e \)), and heavy-lepton flavor neutrinos (\( \nu_x \)). The theoretical distributions for these neutrinos are shown by the blue, orange, and green curves, respectively. Sampled data points are represented by blue circles for \( \nu_e \), orange squares for \( \bar{\nu}_e \), and green triangles for \( \nu_x \).}
    \label{neutrino_spectra_sampled}
\end{figure}

Following the approach of \citep{PhysRevD.109.083019}, we consider only the ratios of average energies and luminosities as relevant parameters. For the energy-integrated flux factors, we adopt $F_{\nu_e}=0.5$, $F_{\bar{\nu}_e}=0.7$, and $F_{\nu_x}=0.8$. The energy-dependent flux factor is modeled as given in \autoref{energy}, assuming 14 energy bins and a spectral cutoff at $E_{\nu,i} \gtrsim 60\,\mathrm{MeV}$. This decreasing trend reflects the expected suppression of flux at higher energies due to the nonlinear dependence of neutrino–matter cross sections in the CCSNe environment. While this treatment is phenomenological, it captures the essential energy dependence of the flux factor. \autoref{neutrino_spectra_sampled} displays the initial spectra consistent with these assumptions before the onset of FFCs.

Given our models, the final neutrino number density in each energy bin is computed as

\begin{equation}
n^{\text{fin}}_{\nu_\beta (\bar{\nu}_\beta), i} = \frac{N^{\text{ini}}_{\nu(\bar{\nu}), i}}{N^{\text{pred}}_{\nu(\bar{\nu}), i}} \, n^{\text{pred}}_{\nu_\beta(\bar{\nu}_\beta), i}, 
\end{equation}
where \( n^{\text{pred}}_{\nu_\beta(\bar{\nu}_\beta), i} \) is the model-predicted number density for neutrino species \( \beta \) in bin \( i \). \( N^{\text{pred}}_{\nu(\bar{\nu}), i} \) and \( N^{\text{ini}}_{\nu(\bar{\nu}), i} \) denote the predicted and initial total (anti)neutrino number densities, respectively.

% These predictions rely solely on the first two radial moments of neutrinos in each energy bin, as typically provided in advanced CCSN and NSM simulations. Our KANs yield errors as low as $\lesssim x\%$ for the electron-channel neutrino number and $\lesssim x\%$ for the relative absolute error in the neutrino moments.

% Here, the spectral error function is defined as:

% \[
% \delta_\nu = \sum_i \frac{n_{\nu,i}}{n_\nu} \left| \frac{\Delta n_{\nu,i}}{n_{\nu,i}} \right|,
% \]
% where \( n_{\nu,i} \) is the neutrino number in the \( i \)-th energy bin, \( \Delta n_{\nu,i} = n_{\nu,i}^{\text{pred}} - n_{\nu,i}^{\text{true}} \), and \( n_\nu = \sum_i n_{\nu,i} \) is the total neutrino number. This definition provides a weighted average of the relative errors across all energy bins. For the results presented in the \autoref{reslut}, we observed \( \delta_{\nu_e} = 0.6 \), \( \delta_{\bar{\nu}_e} = 0.4 \), \( \delta_{\nu_x} = 0.3 \), and \( \delta_{\bar{\nu}_x} = 0.4 \).
% 对比最好的kan，拟合的式子，加上pysindy，评估三者的优缺点
% 可以把符号化的kan，参数化kan，pysindy和真实值放一张图上
To evaluate the accuracy of the reconstructed neutrino energy spectra after FFCs, we introduce the spectral shape error, $\delta_{\nu}$, as our primary metric. This error is calculated as the sum of absolute differences across all energy bins, normalized by the total count of true neutrinos:
\begin{equation}
\delta_\nu = \sum_i \frac{n_{\nu,i}}{n_\nu} \left| \frac{\Delta n_{\nu,i}}{n_{\nu,i}} \right|,
\end{equation}
where \( n_{\nu,i} \) is the neutrino number in the \( i \)-th energy bin, \( \Delta n_{\nu,i} = n_{\nu,i}^{\text{pred}} - n_{\nu,i}^{\text{true}} \), and \( n_\nu = \sum_i n_{\nu,i} \) is the total neutrino number. This definition provides a weighted average of the relative errors across all energy bins. For the results shown in \autoref{fig:all_spectra_combined}, the errors for each model are listed in \autoref{tab:comparison}. Overall, the figure and the data in the table illustrate the ability of the models to reproduce the characteristic spectral shape and peak location across all neutrino species.

To systematically evaluate the performance of different data-driven methods, we conducted a  comparison of five models based on two primary criteria: predictive accuracy and model interpretability. As shown in \autoref{tab:comparison}, our evaluation spans a range of techniques used in this work, from traditional “black-box” neural networks to symbolic regression methods designed for discovering physical laws.

The results illustrate that no single model achieves uniformly superior predictive accuracy across all neutrino species. The Parametric Kolmogorov-Arnold Network (Param. KAN) demonstrates the lowest prediction errors for the $\nu_e$ and $\nu_x$ species, while the baseline MLP model achieves the smallest errors for $\bar{\nu}_e$ and $\bar{\nu}_x$. \textcolor{black}{The Symbolic KAN yields moderate errors overall, performing better than the SINDy-based methods. In contrast, the SINDy and SINDy-SA models, despite their advantages in interpretability—with SINDy-SA being the most interpretable—consistently show the highest prediction errors in this comparison.}

From the perspective of model architecture and interpretability, these methods span a spectrum from “black-box” to “white-box.” The MLP is a classic black-box model, with opaque internal logic. \textcolor{black}{At the other extreme, the SINDy methods serve as white-box models, distilling system dynamics into concise ordinary differential equations (ODEs) that are ideal for uncovering physical laws. Notably, SINDy-SA is specifically designed to deliver the most interpretable, concise expressions.} The KAN model occupies an intermediate position: its parametric form achieves high-fidelity predictions, while its symbolic form can generate interpretable formulas. It is important to note, however, that these symbolic formulas are often highly complex.

% Taken together, our analysis highlights a fundamental trade-off in data-driven scientific modeling: balancing predictive accuracy with physical interpretability. If the primary goal is to achieve the highest possible accuracy in predicting system states, models like the Param. KAN or MLP are advantageous. \textcolor{black}{Conversely, for research focused on discovering and understanding fundamental physical laws, methods such as SINDy and SINDy-SA—which produce concise, interpretable governing equations—are indispensable, even if this comes at the cost of reduced predictive accuracy.} While Symbolic KAN offers a potential pathway for unifying high accuracy with interpretability, its resulting complexity suggests that truly interpretable models must also remain sufficiently concise to yield human-understandable, meaningful physical insight.

\textcolor{black}{Taken together, our analysis highlights a fundamental trade-off in data-driven scientific modeling: balancing predictive accuracy with physical interpretability. Our results exemplify this: if the primary goal is to achieve the highest possible accuracy in predicting system states, models like the Param. KAN or MLP are advantageous. Conversely, for research focused on discovering and understanding fundamental physical laws, methods such as SINDy and SINDy-SA—which produce concise, interpretable governing equations—are indispensable, even if this comes at the cost of reduced predictive accuracy.} \textcolor{black}{What scientists need to do is not a simple compromise, but a strategic choice from a collection of high-performing models. While Symbolic KAN offers a potential pathway for unifying high accuracy with interpretability, its resulting complexity suggests that truly interpretable models must also remain sufficiently concise to yield human-understandable, meaningful physical insight.}

\begin{widetext}
% 上面要去掉i

% --- 无需任何额外宏包，此方案在标准 revtex4-2 模板中测试通过 ---

% --- 使用标准的 table* 环境，标题会自动居中 ---
\begin{table*}[]
\caption{\raggedright
    \label{tab:comparison}
    Comparison of different models for reconstructing post-FFC supernova neutrino energy spectra. 
    The models are evaluated on their reconstruction accuracy, quantified by the spectral shape error $\delta_{\nu}$ for each neutrino ($\nu_e$, $\bar{\nu}_e$, $\nu_x$, $\bar{\nu}_x$). 
    For each method, the values presented correspond to the best-performing result obtained from multiple training runs or configurations. 
    Lower $\delta_{\nu}$ indicates higher accuracy. 
    We also report an overall average accuracy, defined as $(1 - \overline{\delta_{\nu}}) \times 100\%$, where $\overline{\delta_{\nu}}$ is the mean of the $\delta_{\nu}$ values across the four neutrino types.
    Additionally, model form and interpretability are also described.
    \textcolor{black}{
    We use \texttt{SymPy}’s \texttt{count\_ops} function to compute the complexity metric; by counting operations across nested subexpressions, it captures both operation quantity and effective depth, providing a structural complexity indicator.
    }
    \textcolor{black}{
    Parame. KAN denotes the best-performing test result of the original KAN with learned B-spline activations (no symbolic regression).
    Symbolic KAN denotes the symbolic results obtained by applying the symbolic regression procedure (mentioned at the end of \autoref{Construction of Kolmogorov-Arnold Networks}) to the trained KAN.
    }
}
\begin{ruledtabular}
    \renewcommand{\arraystretch}{1.25}
% Using 'c' for centered numerical columns and 'l' for left-aligned text columns.
\begin{tabular}{l ccccc l}
\textbf{Model} & 
\textbf{$\delta_{\nu_e}$} & 
\textbf{$\delta_{\bar{\nu}_e}$} & 
\textbf{$\delta_{\nu_x}$} & 
\textbf{$\delta_{\bar{\nu}_x}$} &
\textbf{Average Accuracy} &
\textbf{Model Form \& Complexity} \\
\hline
% Data is illustrative, based on the hierarchy described in your manuscript.
% I've used the values from your text for Symbolic KAN as an example.
MLP \citep{PhysRevD.109.083019}  & 0.20 & 0.06 & 0.23 & 0.05 &  86\% & Black-box neural network (Too many)\\
Param. KAN        & 0.11 & 0.08 & 0.13 & 0.10 & 90\% & Parametric B-splines (Too many) \\
Symbolic KAN      & 0.20 & 0.18 & 0.13 & 0.16 & 83\% & Complex symbolic formula (1624) \\
\textcolor{black}{SINDy}           & 0.21 & 0.21 & 0.19 & 0.19 & 80\% & Concise ODEs (456) \\
\textcolor{black}{SINDy-SA}           & 0.23 & 0.21 & 0.19 & 0.19 & 80\% & Concise ODEs (295) \\
\end{tabular}
\end{ruledtabular}
\end{table*}

% --- 创建一个横跨双栏的浮动体环境 ---
\begin{figure*}[t!]
    \centering % 将内部所有内容居中

    % --- 第一行图片 ---
    \includegraphics[width=0.48\textwidth]{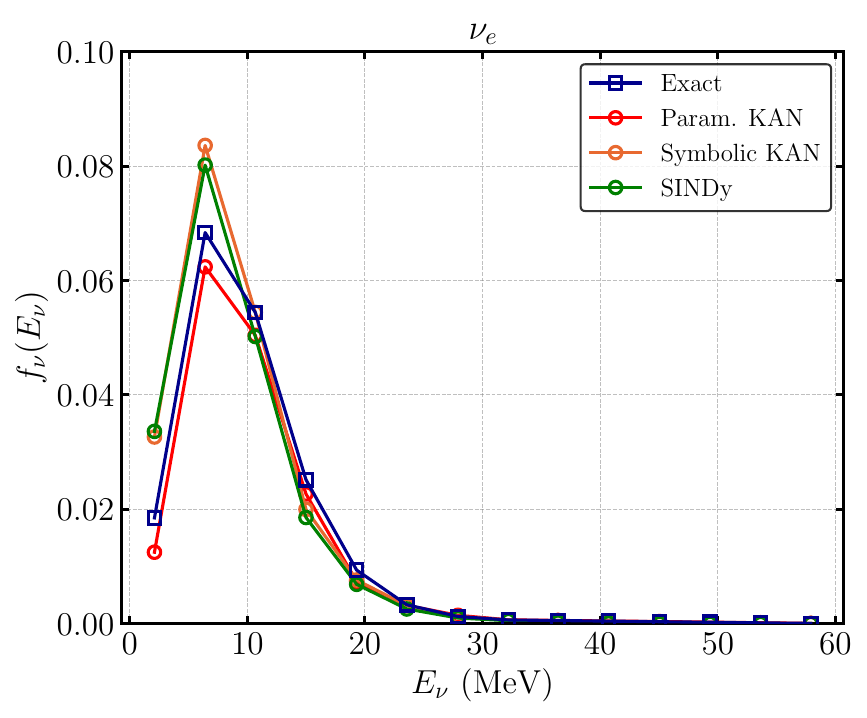}
    \hfill % 在两张图片之间创建弹性水平间距
    \includegraphics[width=0.48\textwidth]{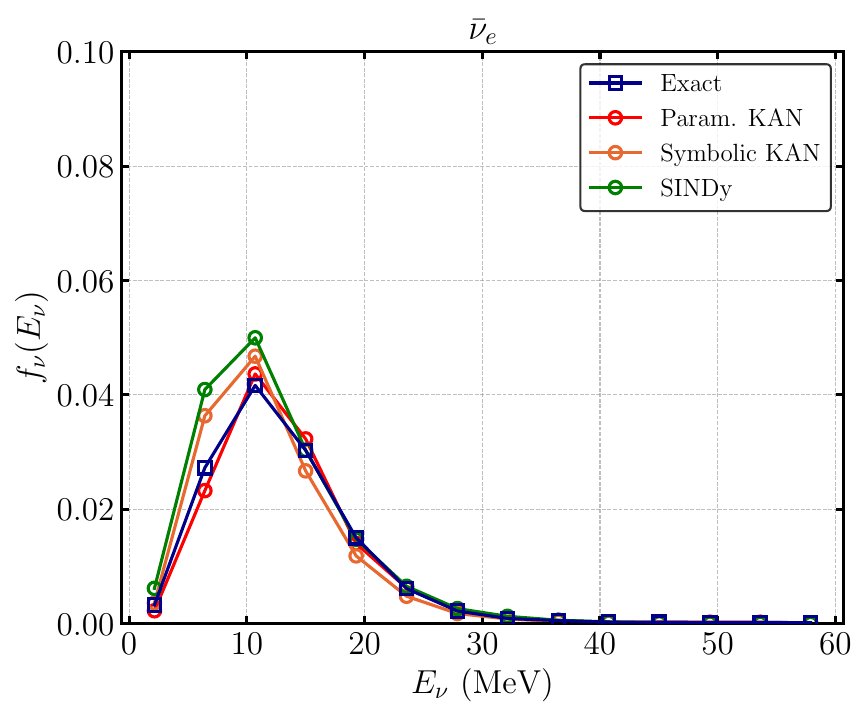}

    \vspace{5mm} % 可选：在两行图片之间增加一点垂直间距

    % --- 第二行图片 ---
    \includegraphics[width=0.48\textwidth]{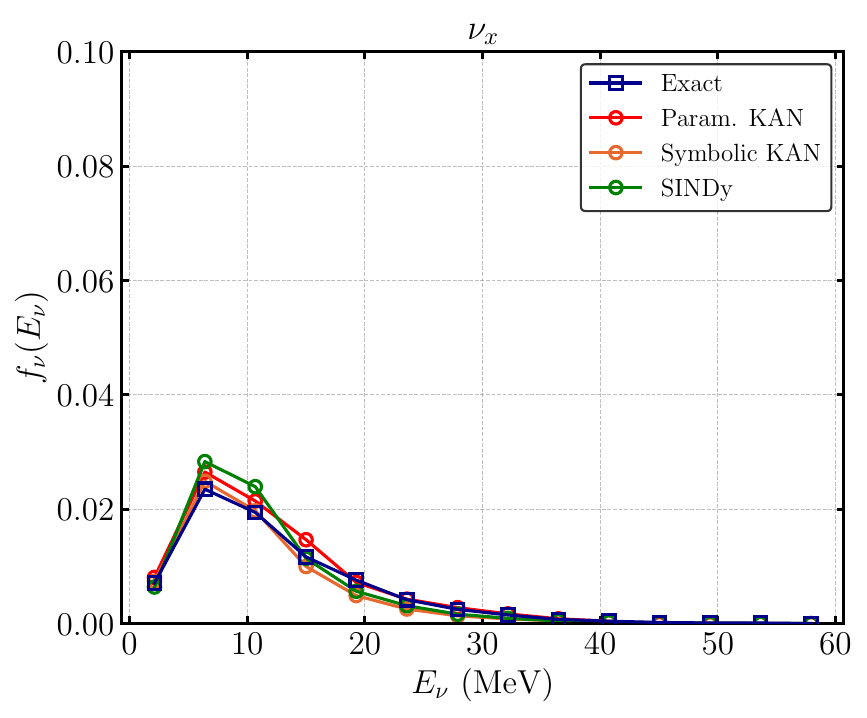}
    \hfill % 同样，创建弹性水平间距
    \includegraphics[width=0.48\textwidth]{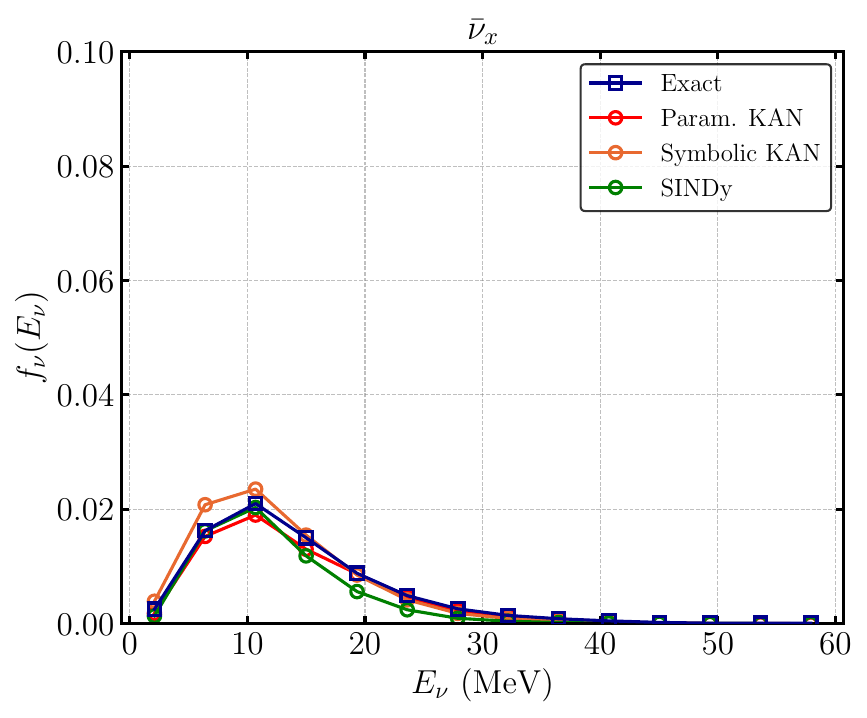}

    % --- 整个大图的总标题和总标签 ---
    \caption{\raggedright
        This figure compares the reconstructed post-FFC energy spectra from our models with exact calculations. The exact calculations are performed using the benchmark setup described in \citep{PhysRevD.109.083019}. The panels show results for $\nu_e$ (top-left), $\bar{\nu}_e$ (top-right), $\nu_x$ (bottom-left), and $\bar{\nu}_x$ (bottom-right). The legend indicates results from Exact (blue squares), Param. KAN (red circles), Symbolic KAN (orange circles), and \textcolor{black}{SINDy} (green circles). 
        % This figure provides a direct visual comparison of the post-FFC energy spectra reconstructed by our different interpretable models—Parametric KAN (Param. KAN), Symbolic KAN, and PySINDy—against the exact solution. 
        % The exact calculation is performed using the benchmark setup described in \citep{PhysRevD.109.083019}. 
        % The four panels display the results for (a) $\nu_e$ (top-left), (b) $\overline{\nu}_e$ (top-right), (c) $\nu_x$ (bottom-left), and (d) $\overline{\nu}_x$ (bottom-right). 
        % These plots visually illustrate the performance trade-offs detailed quantitatively in Table II.
        % Reconstructed post-FFC energy spectra for all relevant neutrino flavors. 
        % The top-left panel corresponds to $\nu_e$, top-right to $\bar{\nu}_e$, 
        % bottom-left to $\nu_x$, and bottom-right to $\bar{\nu}_x$.
        % Each panel compares the exact calculation with predictions from our models.
    }
    \label{fig:all_spectra_combined}
\end{figure*}

\end{widetext}

% \begin{figure*}[!t]
%     \centering
%     % First subplot
%     \subfloat[$\nu_{e}$]{
%         \includegraphics[scale=0.32]{result_e0412 .pdf}
%         \label{re}
%     }
%     % Second subplot
%     \subfloat[$\bar{\nu}_{e}$]{
%         \includegraphics[scale=0.32]{result_eb0412 .pdf}
%         \label{reb}
%     }
%         \subfloat[$\nu_{x}$, $\bar{\nu}_{x}$]{
%         \includegraphics[scale=0.32]{result_nux_nuxb_final0412 .pdf}
%         \label{rec}
%     }
%         \caption{Post-FFC neutrino spectra for $\nu_e$, $\bar{\nu}_e$, and $\nu_x$ ($\bar{\nu}_x$) shown in \autoref{re}, \autoref{reb} and \autoref{rec}, respectively. 
%         Results are obtained using both the KAN approach and the exact method, assuming full knowledge of the neutrino angular distributions.
%         The results from the exact method are consistent with those in \citep{PhysRevD.109.083019}.}
%     \label{reslut}
% \end{figure*}

\section{Data-driven Exploration of Physical Information}\label{Extracting Physical Information from KAN}

\begin{figure}[h]
    \centering % 将图片在栏内居中
    \includegraphics[width=\columnwidth]{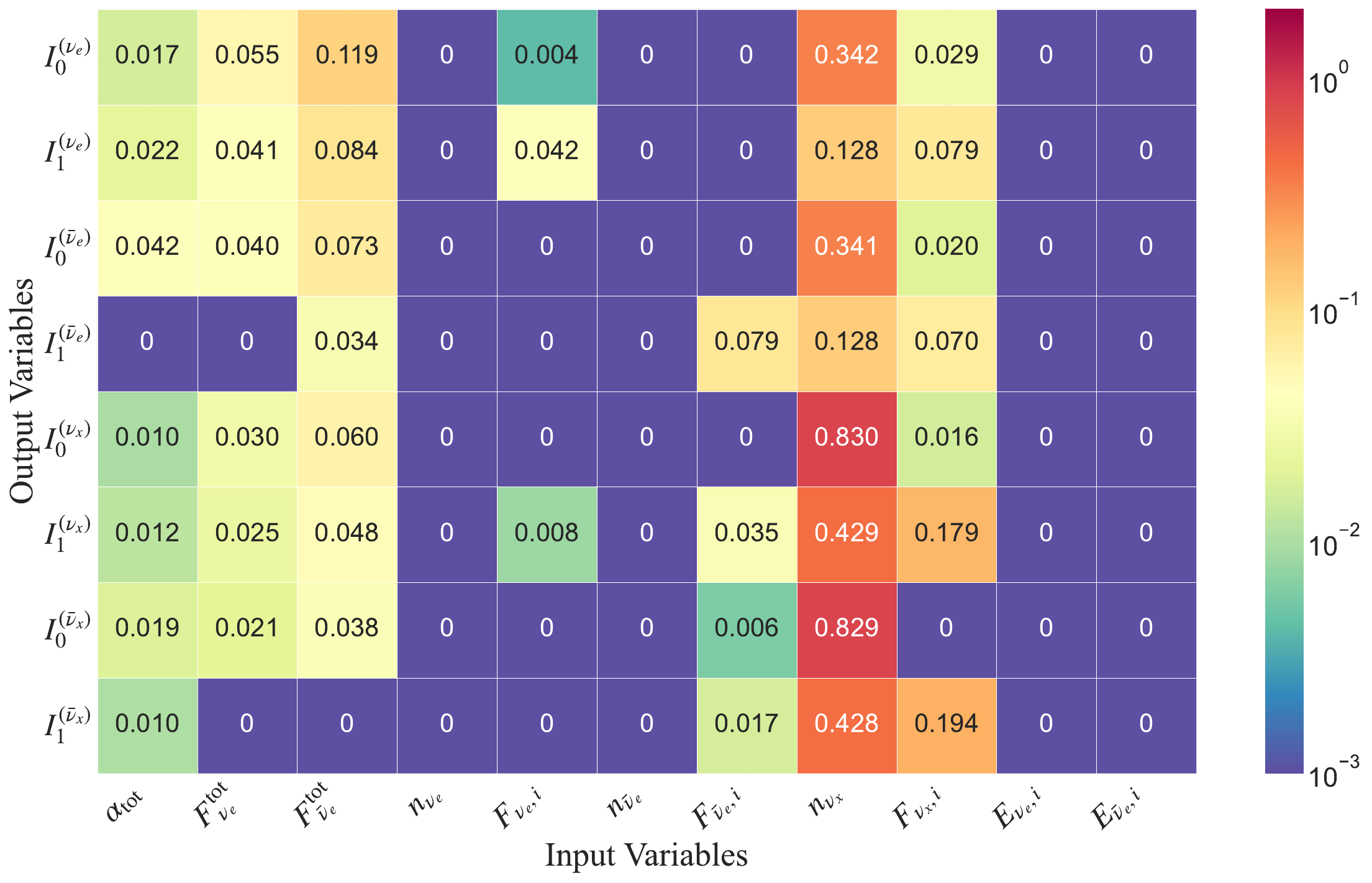}
    \caption{\raggedright
    \textcolor{black}{
    Heatmap of normalized absolute partial derivatives for the SINDy-SA. 
    Each entry reports the sample mean of $|\partial y_j / \partial x_i|$ evaluated at the original data points 
    Warmer colors indicate stronger influence of input variable $x_i$ on output $y_j$, cooler colors indicate weaker influence. 
    \textcolor{black}{Here, we set all values less than $10^{-3}$ to zero.}
    }
    }
    \label{P_heatmap}
    
\end{figure}

\begin{figure}[h]
    \centering % 将图片在栏内居中
    \includegraphics[width=\columnwidth]{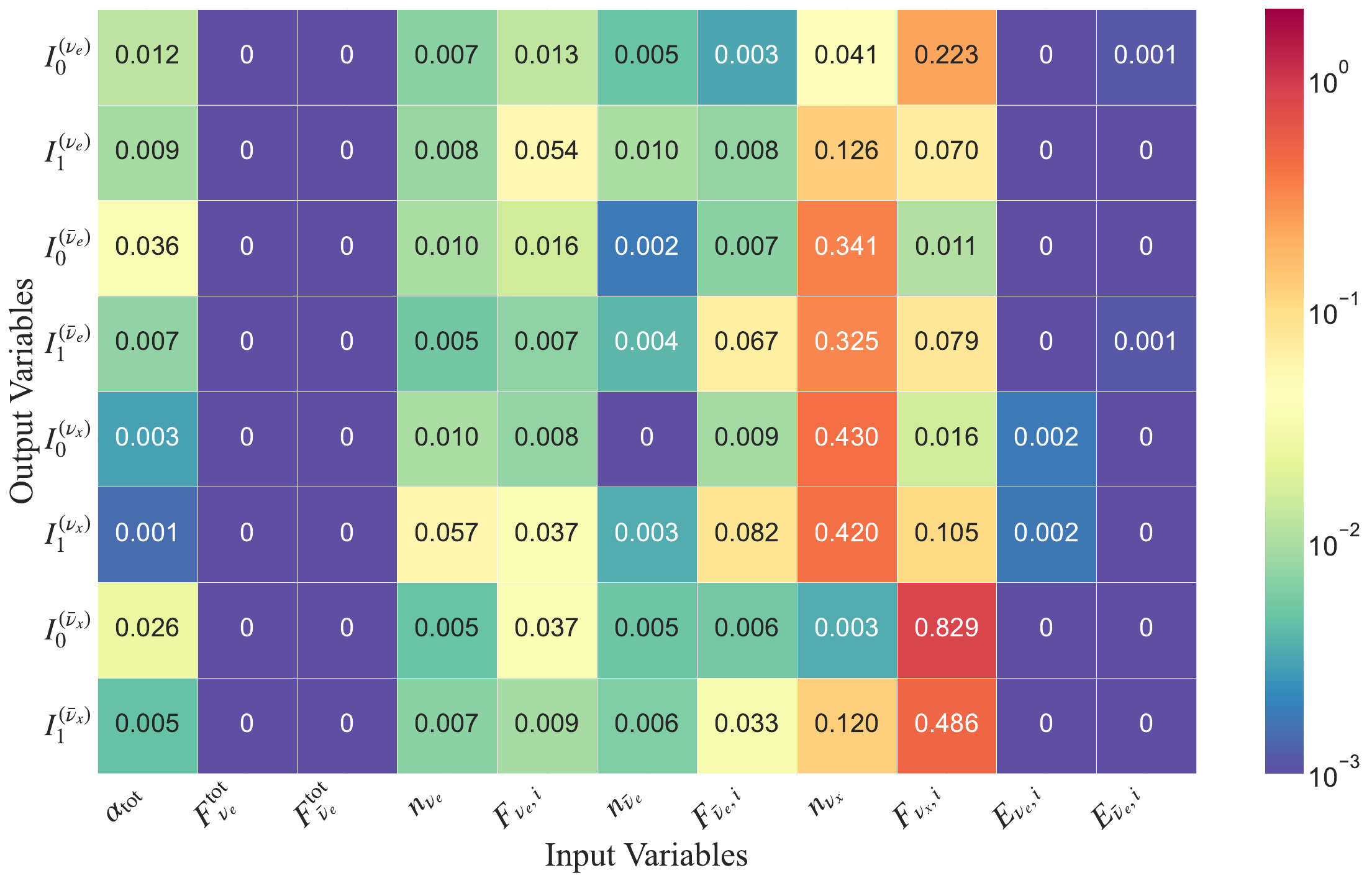}
    \caption{\raggedright
    \textcolor{black}{
    Heatmap of normalized absolute partial derivatives for the symbolic KAN. 
    Each entry reports the sample mean of $|\partial y_j / \partial x_i|$ evaluated at the original data points 
    Warmer colors indicate stronger influence of input variable $x_i$ on output $y_j$, cooler colors indicate weaker influence.
    \textcolor{black}{Here, we set all values less than $10^{-3}$ to zero.}
    }
    }
    \label{P_heatmap_kan}
    
\end{figure}

\textcolor{black}{
Compared with conventional data-driven neural networks, KAN and SINDy provide symbolic models rather than black-box predictors. This enables direct examination of the relationship between inputs and outputs, that is, how each input variable affects the predicted results.
A practical way to study this relationship is to compute partial derivatives. This allows us to see, in an intuitive manner, how every output responds to each input. 
The same idea has been used in symbolic KAN-based convolutional neural networks \citep{huang2025interpretable}. 
In our analysis, we compute the partial derivative of each output with respect to each input, then evaluate these derivatives at the original data points, and take absolute values. \autoref{tab:comparison} shows that the SINDy and SINDy-SA models have almost identical predictive accuracy. However, because SINDy-SA is specifically designed to yield the most simplified form, it is the ideal candidate for our analysis. Therefore, this section will focus solely on the insights gained from the KAN and SINDy-SA models. The resulting heat map is shown in \autoref{P_heatmap} and \autoref{P_heatmap_kan}. We interpret the normalized magnitude as the contribution of each variable to the outcome, where larger values indicate stronger contributions.
}

% 这里下面可以开始看图说话，不过我感觉要改的也不多

\textcolor{black}{\textcolor{black}{As shown in \autoref{kan} and \autoref{P_heatmap_kan}, despite employing diverse machine learning training methodologies, our converged results consistently indicate that the parameter $F_{\text{total}}$ is not utilized in predicting the final neutrino momentum.}} However, satisfactory outcomes cannot be achieved if it's removed from the dataset prior to training. \textcolor{black}{This suggests that $F_{\text{total}}$ is still involved in the training process; a possible explanation is that its spectral dependence likely approximates a constant relationship rather than a variable one, leading to its simplification by the network in the final stages.} This aligns with the previously proposed theory that in a multi-energy neutrino gas, when the FFC growth rate significantly surpasses that of the vacuum Hamiltonian, all neutrinos exhibit a common survival probability determined by the energy-integrated spectrum  (regardless of energy) \citep{PhysRevD.109.083019}.  \textcolor{black}{However, \autoref{kan} shows that the energy input \textcolor{black}{(\(E_{\nu_e,i}\) and \(E_{\bar{\nu}_e,i}\))} still has a weak influence on the predictions of the output. This is consistent with the statement in \citep{PhysRevD.109.083019} that "the survival probabilities are nearly independent of the neutrino energy." While this does not confirm complete energy independence, it implies that any existing dependence is weak and non-dominant, potentially stemming from subtle, underlying physical processes. \textcolor{black}{One explanation is that the growth rate of FFC is too rapid compared to the typical timescale of the vacuum term, making the explicit energy dependence in the vacuum term negligible. In this sense, FFC is almost independent of energy, though not entirely. In particular, when energy-independent flavor conversion from fast flavor instability reaches nonlinear saturation, the energy contribution from the vacuum term could appear, including slow flavor conversion.}} \textcolor{black}{For SINDy-SA, since it is inherently an effective model that seeks low rank, its ability to capture details is lower than that of KAN, which is why its prediction accuracy is lower than that of KAN. However, it can still effectively capture the dominant components in the mapping relationship. From \autoref{P_heatmap}, it can also be observed that the dominant parameters for the predicted output are $n_{\nu_{x}}$ and $F_{\nu_{x},i}$, which is consistent with the conclusions we obtained from KAN.} Utilizing the methodology described above, this study obtains the analytical and semi-analytical dependence of the energy spectrum on its moments. Formally, the dependencies derived from both KAN and SINDy can be represented by matrix equations of an identical form. This may indicate that for most learning tasks, there exists an underlying, well-performing, low-rank neural network solution \citep{li2025approximation}.

% Although KANs can yield symbolic analytical expressions, they remain considerably more complex than those obtained via SINDy. Therefore, in this section, we primarily focus on what insights can be extracted from the SINDy and SINDy-SA results. As shown in \autoref{tab:comparison}, SINDy and SINDy-SA produce nearly identical prediction results. Therefore, we choose to use the simplified form obtained from SINDy-SA for further analysis.

%While KAN can yield symbolic expressions, their complexity makes them less suitable for extracting physical insights than the more concise SINDy-based methods. 

\textcolor{black}{Across all input parameters, the input related to $\nu_x$ ($n_{\nu_x}$ and $F_{\nu_{x},i}$) is observed to be the most significant and exhibits the strongest correlation with all outputs. Especially, $n_{\nu_x}$ consistently holds a prominent position across different models; this is not a surprising result.} \textcolor{black}{Because it has a similar structure to previous work showing that the frequency of collective oscillations scales as $\sim \mu_{\nu}$ \citep{Airen:2018nvp}, where $\mu_{\nu}=\sqrt{2} G_F n_\nu$ represents the potential due to a neutrino density $n_\nu$.} \textcolor{black}{\textcolor{black}{Furthermore, because \(F_{\nu,i}\) is defined as \(F_{\nu,i}=(I_1/I_0)_{\nu,i}\) with \(I_{0,i}=n_{\nu,i}\), \(F_{\nu_x,i}\) has an explicit dependence on \(n_{\nu_x}\). This dependence could help account for the relatively large weight of \(F_{\nu_x,i}\) in the heat map. Still, this should be read as a correlation rather than proof of causal dominance, since \(F_{\nu_x,i}\) also reflects \(I_{1,\nu_x,i}\) and model-specific interactions among inputs.} Our analysis indicates that reactions affecting the initial number density of heavy-lepton neutrinos, such as electron-positron pair annihilation, play a more pivotal role in shaping the outcome of FFCs within our model.This is because the system's evolution exhibits the highest sensitivity to the initial number density of heavy-lepton neutrinos, far exceeding its sensitivity to the electron neutrino density or other physical parameters. \textcolor{black}{From another perspective, through flavor conversion, the energy spectra of heavy-leptonic flavors are converted to the electron-type. In this sense, it seems normal that $\nu_x$ makes a significant contribution. By contrast, it is strange that the heatmap of $\nu_x$ (output) from $\nu_e$ (input) is small. The potential possibility is that it is related to the hardness ratio of the energy spectra of $\nu_x$. A more detailed discussion and study on this point can be conducted after having data with richer physical details.} }

% \textcolor{black}{One potential explanation for our finding—that $n_{\nu_x}$ has a prominent role in influencing energy spectrum predictions—may be related to the unique angular distribution of $\nu_x$.} The fact that its angular spread is the smallest among all flavors \citep{Tamborra:2017ubu} could provide favorable conditions for the $\nu$ELN crossing required to trigger FFCs, thus offering a possible physical basis for our model's result. \textcolor{black}{ This suggests a possible indirect coupling through energy transport and hydrodynamic conditions: once activated, FFCs might boost the luminosity of heavy-lepton neutrinos, potentially contributing to significant changes in the supernova's energy transport \citep{Ehring:2023lcd,Nagakura:2023mhr}.}

In summary, from the heatmap computed using the KAN and SINDy-SA framework, we find that among the many input parameters determining the asymptotic values of the neutrino radial moments $I_0$ and $I_1$, the initial number density of heavy-lepton neutrinos always exhibits significant correlation. The relevance values for almost all other parameters are negligible. It is important to note that most of these parameters in fact are the initial-state combinations of the neutrino radial moments prior to the onset of FFCs. \textcolor{black}{ An interpretation was previously provided at the beginning of this section using the KAN model analysis, and the partial derivative method used in this section, as well as arrives at the same conclusion. } This consistency may prompt us to consider the problem from a new perspective: while the triggering of FFCs is known to be dependent on the details of the initial angular distribution, the connection between the asymptotic outcome and those specific initial conditions weakens rapidly once FFCs are triggered, and the system's dynamics become dominated by macroscopic properties such as the global particle number density. \textcolor{black}{It is important to note that we do not have data on the entire evolutionary process over time, so these ideas are speculative.}

This behavior shares a conceptual similarity with the well-known \lq \lq sandpile model\rq \rq \ in dynamical systems exhibiting self-organized criticality \citep{1987PhRvL..59..381B}: the triggering of an avalanche may be caused by a single, randomly falling grain of sand, but the size, shape, and duration of the avalanche are almost entirely independent of the properties of that grain, instead being determined by the sandpile's overall \lq \lq critical state\rq \rq \ , such as its slope and density. Once the sandpile reaches this critical state, there is no longer any correlation between the system's response to a perturbation and the details of that perturbation itself. Similarly, in our scenario, specific initial combinations of $I_0$ and $I_1$ act as the triggering grain of sand, serving mainly to determine whether the FFC \lq \lq avalanche\rq \rq \ is initiated. \textcolor{black}{ Once triggered, the system’s nonlinear dynamics quickly "forget" the details of the trigger, and the final asymptotic state is governed by some of the system's global macroscopic properties—in this study, potentially represented by the heavy-lepton neutrino number density because the FCCs' output energy spectrum is most sensitive to its initial density. It should be noted that this has potential consistency with the results of \citep{PhysRevD.109.083019}, while offering a view of the underlying dynamics.}

\section{Discussion}\label{Discussion}

\textcolor{black}{Our investigation reveals a notable trade-off between predictive accuracy and model interpretability in describing post-FFC neutrino dynamics.} As shown in \autoref{tab:comparison}, the parametric KAN sets a new benchmark for accuracy, surpassing previous MLP-based models in total errors. However, this peak performance comes at the cost of interpretability. The process of deriving an explicit symbolic formula from the KAN inevitably introduces inaccuracies. We identify three primary sources for this precision loss: 
(i) the intrinsic difficulty in fitting certain learned B-spline activation functions to elementary functions with high fidelity; 
(ii) our strategy of averaging B-spline parameters across multiple runs to ensure robustness , which means the final model does not represent the single best-performing run; and 
(iii) the fact that even with averaging, some activation functions did not achieve a coefficient of determination ($R^2$) greater than 0.99 during symbolic regression. 

Furthermore, the resulting symbolic KAN formulas, while analytical, are exceedingly complex. Our attempts to reduce this complexity by simplifying the network architecture (e.g., use fewer nodes or layers, prune network) invariably led to a significant degradation in accuracy, highlighting a practical limitation of this approach for our specific problem.

\textcolor{black}{On the other end of the spectrum, SINDy delivers what the symbolic KAN struggles with: a remarkably concise, low-rank model in the form of simple governing equations.} This elegance and simplicity, however, are paid for with lower predictive accuracy compared to the neural network frameworks. We hypothesize this accuracy gap stems from the differing flexibility of the optimization frameworks. Modern neural network architectures, like our KAN, can seamlessly integrate complex, physics-based prior knowledge. For instance, our KAN model successfully incorporated three distinct physics-inspired loss terms via a dynamic weighting mechanism to enforce physical conservation laws. \textcolor{black}{Such intricate, multi-term constraints are not as easily implemented within the standard SINDy workflow, potentially limiting its ability to converge to a more physically precise solution.}

Ultimately, our work demonstrates that there is no single \lq \lq best\rq \rq \ interpretable model for this problem; rather, the optimal choice depends on the scientific objective. For developing high-fidelity surrogate models intended for integration into larger hydrodynamic simulations where precision is paramount, the parametric KAN or MLP is the superior choice. Conversely, for the goal of gaining physical insight, generating new hypotheses, and discovering the simplest possible governing laws of the system, \textcolor{black}{SINDy is unparalleled}, despite its reduced accuracy. This comparative analysis provides a valuable roadmap for researchers applying machine learning in astrophysics, clarifying the distinct advantages of different interpretable techniques. Future work could focus on developing hybrid methods that leverage the strengths of both approaches. For instance, one could use the sparse governing equations from SINDy as a regularization term, or incorporate insights from GSA to guide the learning of a KAN, thereby pursuing the dual goals of accuracy and interpretability simultaneously.

\section{Conclusion}\label{Conclusion}

In this work, we use and compare two interpretable frameworks, KANs and SINDy, as alternatives to 'black-box' models like MLPs for FFCs in supernova neutrinos.
Based on these models, we predict post-FFC spectra using only the first two radial moments from modern CCSNe and NSM simulations, which are typically available in large-scale simulations. 
This provides a pathway to embed FFC physics into these simulations via surrogate models that are not only accurate but also more physically transparent. 
Our results suggest that, for such complex physical systems, there can be a trade-off between prediction accuracy and model interpretability.
% Our analysis shows that the parametric KAN offers the highest prediction accuracy and serves as an excellent alternative to standard MLP models. However, when expressed in symbolic form through regression, the KAN model sacrifices some accuracy and becomes more complex, which can limit its direct use for physical insight. \textcolor{black}{In contrast, SINDy performs well where symbolic KAN struggles.} It produces a simpler and more elegant low-rank dynamical model, though this simplicity comes at the cost of lower predictive accuracy compared to neural network methods. 
Despite these trade-offs, both interpretable frameworks successfully identify the same key physical feature: the relatively weak dependence on energy as well as on $F_{\nu_e}^{\mathrm{tot}}$ and $F_{\bar{\nu}_e}^{\mathrm{tot}}$ in the post-FFC neutrino first two radial moments. Moreover, the derivatives-based contribution heatmaps of KAN and SINDy results highlight that the initial number density of heavy-lepton neutrinos is the main factor influencing FFCs evolution results. 
% \textcolor{pink}{These interpretable findings demonstrate the benefits of interpretability and underscore the value of moving beyond black-box models.}
\textcolor{black}{These interpretable findings highlight the benefits of interpretability and demonstrate the value of models with more transparent structures.}

In summary, our research enhances the effectiveness of NN in predicting the asymptotic outcomes of FFCs in a multienergy neutrino gas. Leveraging this identified structure, the post-conversion neutrino energy spectra are reconstructed with an average accuracy of up to 90\%. Furthermore, we employ interpretable models to analyze key factors influencing the evolution during FFCs. \textcolor{black}{Future work should focus on extending the study to more realistic neutrino gases with richer physical details.} Interestingly, recent work provides a systematic treatment of inhomogeneous many-body (MB) dynamics of the fast flavor instability, demonstrating that MB effects are expected to persist even in thermodynamically large neutrino ensembles \citep{2025Laraib}. These advancements will improve the feasibility of incorporating FFCs into CCSNe and NSM simulations, thereby enhancing our ability to accurately model and predict these extreme astrophysical phenomena. 

This study also aims to establish a methodological framework for interpretable machine learning that supports genuine data-driven physical discovery in astronomy and astrophysics, going beyond mere prediction. With the advent of astronomical observatories like Rubin \citep{Andreoni:2024pkp}, FAST \citep{2011IJMPD..20..989N}, and JWST \citep{2023arXiv231011912S}, and upcoming facilities, especially next-generation facilities for multi-messenger astronomy such as Roman \citep{Lam:2023qtv}, CSST \citep{Gong:2025ecr}, Taiji \citep{2018arXiv180709495R}, Tianqin \citep{Luo:2025sos}, LISA \citep{amaro2017laser}, and TRIDENT \citep{TRIDENT:2022hql}, the volume of future astronomical data is expected to be immense. In addition, platforms such as Astro-COLIBRI: A Comprehensive Platform for Real-Time Multi-Messenger Astrophysics are being developed as tools to rapidly compile and correlate relevant information for each new event \citep{Schussler:2025eyq}. Consequently, extracting physical laws from this data represents a valuable research area. While current efforts are rudimentary, future work will focus on more clearly articulating human-understandable, data-driven physical laws.

%\clearpage
\begin{acknowledgments}

Thanks to Sajad Abbar for providing the dataset used in this study and for his guidance, as well as to Kaifan Ji, Shun Zhou and Xuwei Zhang for their useful discussions. This work received support from the National Natural Science Foundation of China under grants  12541303, 12288102, 12373038, 12125303, 12090040/3, U2031204 and 12433009 (with Junda Zhou as the sole recipient); the Natural Science Foundation of Xinjiang No. 2022TSYCLJ0006; the science research grants from the China Manned Space Project No. CMS-CSST-2021-A10; the National Key R\&D Program of China No. 2021YFA1600401 and No. 2021YFA1600403; the Natural Science Foundation of Yunnan Province Nos. 202201BC070003 and 202001AW070007; the International Centre of Supernovae, Yunnan Key Laboratory No. 202302AN360001; and the Yunnan Revitalization Talent Support Program $-$ Science \& Technology Champion Project No. 202305AB350003.

\end{acknowledgments}

% Arxiv
% \hspace{2em}
\section*{\textit{Code and Data Availability}}\label{code}
% The source code for the methods implemented in this study will be made publicly available in a public repository upon acceptance of the manuscript. 

% The simulation dataset used for building our models was originally developed and described in \citep{PhysRevD.109.083019}, which provides comprehensive details regarding its parameters and access.
\textcolor{black}{The source code for the interpretable data-driven methods presented in this study, including the implementation of our KAN and SINDy frameworks, is available in an \href{https://anonymous.4open.science/r/interpretable-data-driven-methods-for-FFCs-1827/}{anonymous repository} \footnote{\href{https://anonymous.4open.science/r/interpretable-data-driven-methods-for-FFCs-1827/}{https://anonymous.4open.science/r/interpretable-data-driven-methods-for-FFCs-1827/}}. }

The simulation dataset used for building our models was originally developed and described in \citep{PhysRevD.109.083019}, which provides comprehensive details regarding its parameters and access.

\bibliographystyle{elsarticle-num}
\bibliography{apssamp}% Produces the bibliography via BibTeX.

\end{document}